\documentclass[online]{tlp}

\newtheorem{example}{Example}  

\usepackage{bold-extra}
\usepackage{amsmath}
\usepackage{graphicx}
\usepackage{multirow}
\usepackage{dsfont}
\usepackage{subcaption}
\usepackage{wrapfig} 
\usepackage{comment} 
\usepackage{amssymb}
\usepackage{amsmath}
\usepackage{graphicx}
\usepackage{multirow}
\usepackage{colortbl}
\usepackage{multicol}
\usepackage{xspace}
\usepackage{xcolor}
\usepackage{booktabs}
\definecolor{linkcolor}{HTML}{A93C93}
\usepackage[
textwidth=0.125\textwidth,textsize=scriptsize]{todonotes}
\DeclareCaptionLabelFormat{mysubfig}{Fig.~\thefigure#2:}

\usepackage{listings}
\usepackage{adjustbox}
\usepackage{relsize}
\definecolor{PrologPredicate}{RGB}{0,0,200}
\definecolor{PrologVar}      {RGB}{105,0,175}
\definecolor{PrologComment}  {RGB}{169,082,044}
\definecolor{PrologOther}    {rgb}{0.2,0.2,0.2}
\definecolor{PrologString}   {RGB}{070,120,200}
\usepackage{listings} 
\usepackage{textcomp}
\newcommand{\code}{\lstinline[style=MyInline]}
\newcommand{\codeB}{\lstinline[style=MyInline2]}
\lstdefinestyle{MyInline}
{
  basicstyle = \relsize{-0.5}\ttfamily\color{PrologPredicate},
  breaklines = true,
  breakatwhitespace=true,
  literate =
  {?-}{{?-\,}}3
  {:-}{{:-\,}}3
}
\lstdefinestyle{MyInline2}
{
  basicstyle = \relsize{-0.5}\ttfamily\color{PrologPredicate},
  breaklines = true,
  breakatwhitespace=true,
  moredelim = {*[s][\color{PrologVar}]{(}{)}},
  moredelim = {*[s][\color{PrologString}]{'}{'}},
  moredelim = {*[s][\color{PrologOther}]{:-}{.}},
  literate =
  {?-}{{?-\,}}3
  {:-}{{:-\,}}3
}
\lstdefinestyle{MySCASP}
{
  showlines=true,
  numbersep=1em,    
  xleftmargin=0.35cm,  
  numberstyle=\tiny,
  numbers=left,
  stepnumber=1,
  breaklines = false,
  mathescape = true,
  escapechar = @,
  escapeinside = {-<}{>-},
  keywords = {},
  upquote = true,
  basicstyle = \fontencoding{T1}\ttfamily\strut\color{PrologPredicate}\small\linespread{0.75},  
  basewidth = 0.44em,
  moredelim = {*[s][\color{PrologVar}]{(}{)}},
  moredelim = {*[s][\color{PrologString}]{'}{'}},
  moredelim = {*[s][\color{PrologOther}]{:-}{.}},
  commentstyle = \mdseries\color{PrologComment},
  escapebegin=\color{PrologVar},
  morecomment=[l]\%,
  literate     =
  {|}{{|}}2
  {&(}{{(}}1
  {&)}{{)}}1
  {.=.}{{\,\#=\,}}2
  {.<.}{{\,\#<\,}}2
  {.>.}{{\,\#>\,}}2
  {.<>.}{{\,\#\textdoublebarslash\,}}2
  {.=<.}{{\,\#=<\,}}3
  {.>=.}{{\,\#>=\,}}3
}
\usepackage{tikz}
\newcommand{\linkIcon}{%
    \tikz[x=1.2ex, y=1.2ex, baseline=-0.05ex]{%
        \begin{scope}[x=1ex, y=1ex]
            \clip (-0.1,-0.1) 
                --++ (-0, 1.2) 
                --++ (0.6, 0) 
                --++ (0, -0.6) 
                --++ (0.6, 0) 
                --++ (0, -1);
            \path[draw, 
                line width = 0.5, 
                rounded corners=0.5] 
                (0,0) rectangle (1,1);
        \end{scope}
        \path[draw, line width = 0.5] (0.5, 0.5) 
            -- (1, 1);
        \path[draw, line width = 0.5] (0.6, 1) 
            -- (1, 1) -- (1, 0.6);
        }%
    }

\newcommand{\gmidrule}{\arrayrulecolor{gray}\specialrule{0.1pt}{1pt}{2pt}\arrayrulecolor{black}}

\newcommand{\sfigCapVspaceB}{\vspace{-0.8em}}      
\newcommand{\figVspaceB}{\vspace{-1.5em}}          

\newcommand{\secVspaceB}{}       
\newcommand{\secVspaceA}{}       

\newcommand{\ssecVspaceB}{}      
\newcommand{\ssecVspaceA}{}      

\newcommand{\sssecVspaceB}{}     
\newcommand{\sssecVspaceA}{}     

\newcommand{\parVspaceB}{}       

\newcommand{\artifactDoi}{\raisebox{-0.2em}{\href{https://doi.org/10.5281/zenodo.19217875}{%
\includegraphics[height=1em]{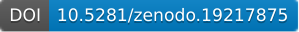}%
}}}

\newcommand{\clingolpx}{clingo-lpx\xspace}
\newcommand{\clingcon} {clingcon\xspace}
\newcommand{\clingo}   {clingo\xspace}
\newcommand{\scasp}    {s(CASP)\xspace}

\newcommand{\EC}[1][]{\texttt{EC#1}\xspace}
\newcommand{\DEC}[1][]{\texttt{DEC#1}\xspace}
\newcommand{\HEC}[1][]{\texttt{HEC#1}\xspace}

\newcommand{\repo}{\href{https://github.com/ovasicek/hybrid-ec}{https://github.com/ovasicek/hybrid-ec}}
\newcommand{\github}{\href{https://github.com/ovasicek/hybrid-ec}{github}}
\newcommand{\linkByIcon}[1]{\href{https://github.com/ovasicek/hybrid-ec/blob/master/#1}{\linkIcon{}}}
\newcommand{\linkHecExample}[2]{\href{https://github.com/ovasicek/hybrid-ec/blob/master/hybrid-clingo/examples/#1}{#2}}
\newcommand{\linkHecExampleECASP}[2]{\href{https://github.com/ovasicek/hybrid-ec/blob/master/hybrid-clingo/examples-ec2asp/#1}{EC2ASP-#2}}

\newcommand{\linkIconHecAx}{\href{https://github.com/ovasicek/hybrid-ec/blob/master/hybrid-clingo/axioms/hec.lp}{\linkIcon{}}}
\newcommand{\linkIconDecAx}{\href{https://github.com/ovasicek/hybrid-ec/blob/master/clingo/axioms/dec_clingo.lp}{\linkIcon{}}}

\newcommand{\linkIconHecEx}[1]{\href{https://github.com/ovasicek/hybrid-ec/blob/master/hybrid-clingo/examples/#1}{\linkIcon{}}}
\newcommand{\linkIconHecExECASP}[1]{\href{https://github.com/ovasicek/hybrid-ec/blob/master/hybrid-clingo/examples-ec2asp/#1}{\linkIcon{}}}
\newcommand{\linkIconDecEx}[1]{\href{https://github.com/ovasicek/hybrid-ec/blob/master/clingo/examples/#1}{\linkIcon{}}}
\newcommand{\linkIconDecExECASP}[1]{\href{https://github.com/ovasicek/hybrid-ec/blob/master/clingo/examples-ec2asp/#1}{\linkIcon{}}}

\begin{document}

\lefttitle{\sn{Va\v{s}\'{i}\v{c}ek} et al.}

\jnlPage{1}{8}
\jnlDoiYr{2021}
\doival{10.1017/xxxxx}

\title[Event Calculus Meets Hybrid ASP]{%
Event Calculus Meets Hybrid ASP%
}

\begin{authgrp}
    \author{\orcidau[0000-0002-4944-2198]{\sn{Va\v{s}\'{i}\v{c}ek}, \gn{Ond\v{r}ej}}}
    \affiliation{Faculty of Information Technology, Brno University of Technology, Brno, Czech Republic}
    \email{ivasicek@fit.vut.cz}
    \author{\orcidau[0000-0003-4148-311X]{\sn{Arias}, \gn{Joaqu\'{i}n}}}
    \affiliation{CETINIA, Universidad Rey Juan Carlos, M\'{o}stoles, Spain}
    \email{joaquin.arias@urjc.es}
    \author{\orcidau[0000-0002-4828-6728]{\sn{Fiedor}, \gn{Jan}}}
    \affiliation{Faculty of Information Technology, Brno University of Technology, Brno, Czech Republic}
    \affiliation{Honeywell International S.R.O., Brno, Czech Republic}
    \email{ifiedor@fit.vut.cz}
    \author{\orcidau[0000-0001-9727-0362]{\sn{Gupta}, \gn{Gopal}}}
    \affiliation{Computer Science Department, UT Dallas, Richardson, TX, USA}
    \email{gupta@utdallas.edu}
    \author{\orcidau[0000-0001-9572-1799]{\sn{K\v{r}ena}, \gn{Bohuslav}}} 
    \affiliation{Faculty of Information Technology, Brno University of Technology, Brno, Czech Republic}
    \email{krena@fit.vut.cz}
    \author{\orcidau[0009-0003-1669-5353]{\sn{N\v{e}mec}, \gn{Jakub}}}
    \affiliation{Faculty of Informatics, Masaryk University, Brno, Czech Republic}
    \email{jaknem@mail.muni.cz}
    \author{\orcidau[0000-0001-5546-9939]{\sn{Romero}, \gn{Javier}}}
    \affiliation{Institute of Computer Science, University of Potsdam, Potsdam, Germany}
    \email{javier.romero.davila@uni-potsdam.de}
    \author{\orcidau[0000-0002-2746-8792]{\sn{Vojnar}, \gn{Tom\'{a}\v{s}}}}
    \affiliation{Faculty of Informatics, Masaryk University, Brno, Czech Republic}
    \affiliation{Faculty of Information Technology, Brno University of Technology, Brno, Czech Republic}
    \email{vojnar@fi.muni.cz}
\end{authgrp}

\history{ %
\sub{7 February 2026;} %
\acc{4 May 2026} %
}

\maketitle

\begin{abstract}
Event Calculus (EC) implemented in answer set programming (ASP) has proven suitable for specifying requirements on safety-critical systems thanks to its elegant representation of both discrete and continuous changes and its semantic closeness to semi-formal natural language. However, continuous changes and the size of value domains of time and system properties (fluents) pose significant challenges. Grounding-based ASP solvers, e.g., clingo, which implement Discrete EC (DEC), lead to combinatorial explosion in program size and inaccurate representation. The grounding-free s(CASP) does not discretize but struggles with non-termination due to its top-down execution. This paper introduces Hybrid EC, an extended axiomatization of DEC, that tackles the challenges via functional fluents and a mapping of time to abstract steps. We implement it using clingcon and clingo-lpx (Hybrid ASP systems over integers and rationals, respectively) where the value (dense) domains of fluents and time are represented as linear constraints and evaluated by external solvers, while ensuring termination whenever solutions exist. We validate both implementations on a number of examples and observe that they are unaffected by the size of the domains and that handling rationals does not impact scalability. Most importantly, the ability of clingo-lpx to handle dense domains enables accurate modeling of continuous change.
\end{abstract}


\secVspaceB{}\section{Introduction and Motivation}\secVspaceA{}
\label{sec:intro}

Modern cyber-physical systems (CPSs) are rapidly growing in size and complexity, making rigorous verification increasingly costly. Using high-level, near-natural specifications with early verification can help mitigate this, but such requirements must first be translated into an appropriate formal language for verification.

Event Calculus (EC) has recently been advocated as a suitable formalism for this purpose, e.g., by \cite{gupta-train}, \cite{clarissa-assurance}, and \cite{iclp24-pcapump-ec-scasp}. EC is a logical formalism for reasoning about events and change, and is well-suited for the given purpose due to its small semantic gap with semi-formal natural language requirements. In particular, \cite{iclp24-pcapump-ec-scasp} illustrates this by analyzing a real-life safety-critical system with an open-source high-level specification, which led to the discovery of a number of inconsistencies and a safety violation in the specification.

However, while EC allows natural modeling of CPS specifications, the resulting models typically use large or even dense value domains and describe continuous behavior. Analyzing such models is a significant challenge for current state-of-the-art general-purpose reasoning tools for EC, which are based on \emph{answer set programming} (ASP). Grounding-based ASP solvers, e.g., \clingo{} by~\cite{clingo-paper}, require discretization of dense domains, which can lead to inaccurate representation of the system.
This can be avoided on a case-by-case basis by finding the right discretization (e.g., increasing the precision from seconds to milliseconds),
but then the value and time domains can easily become large, making the solvers run out of resources due to an explosion in the size of the grounded program. The grounding-free ASP solver \scasp{} by~\cite{arias18-iclp-scasp} does not discretize but faces non-termination issues.

To address these issues, we introduce \emph{Hybrid EC}, an extension of the Discrete EC axiomatization of \cite{mueller_book-fixed}. Hybrid EC introduces the so-called \emph{functional fluents} to represent various system properties. The values of such fluents are represented using constraints. Moreover, Hybrid EC replaces the integer-time semantics of DEC with \emph{abstract step indices} representing the sequence of system evolution. These steps are subsequently mapped to time through constraints. 

We then utilize \clingcon{} and \clingolpx, Hybrid ASP solvers supporting linear constraints over integers and rationals, to implement the proposed Hybrid EC through the \HEC{} axioms.
Execution under Hybrid ASP avoids the grounding-size explosion with respect to time and value domains. More importantly, the \HEC{} implementation enables accurate modeling of and reasoning about continuous change since \clingolpx supports the representation of dense domains using rationals. This allows reasoning without discretization and eliminates inaccuracies due to discretized representations.
    
We have implemented and analyzed several examples from the literature to demonstrate the modeling capabilities and performance of our approach using \clingcon{} and \clingolpx{}. We compare the results with those obtained from the plain ASP solver \clingo{}.  Our hybrid approach scales independently of the size of the time and value domains, while the plain ASP approach does not.  Furthermore, the use of rationals enables accurate modeling of continuous change, which was not possible previously, without affecting scalability compared to reasoning over integers.

\secVspaceB{}\section{Background}\secVspaceA{}
\label{sec:background}

This section provides a brief overview of Event Calculus (EC),
a~formalism for reasoning about events and change, initially proposed
by~\cite{kowalski86} and \cite{shanahan97-book-frame_problem}, and
Hybrid ASP, an extension of ASP that incorporates linear constraints.

\ssecVspaceB{}\subsection{Event Calculus}\ssecVspaceA{}
\label{sec:event-calculus}

Several versions of EC have been proposed, see
survey by \cite{mueller-ec-chapter}, among which the
most relevant are the \textit{Event Calculus} (\EC{}) and the \textit{Discrete
Event Calculus} (\DEC{})
by \cite{mueller-ec-through-sat} formulated for non-monotonic reasoning using first order logic and circumscription.
EC was later reformulated by~\cite{reformulating} in the theory of stable models and translated into ASP.
Throughout the paper, we use the ASP encoding of \DEC{} in code listings and as the base of our hybrid formulation of EC.

There are three basic concepts in EC:
\textit{events} representing actions that may occur in the
world;
\textit{fluents} modeling time-varying properties of the
world;
and \textit{timepoints} representing instants of time.
Events may happen at a~timepoint.
Properties represented by fluents change discretely via events and stay constant between occurrences of the events due to inertia.
When released from inertia, they change continuously via \mbox{\textit{(anti-)trajectories}} or are non-deterministic.
Predicates used in \EC{} and \DEC{} are listed in Table~\ref{fig:ec-predicates}.
A comprehensive introduction to EC can be found in~\cite{mueller_book-fixed}.

\begin{table}[tb]
  \setlength{\tabcolsep}{1pt}
  \linespread{0.9}
  \centering
  \small
  \caption{Predicates used in \EC{} and \DEC{}. Based on Table 17.4 from~\cite{mueller-ec-chapter}.} 
  \label{fig:ec-predicates}
  \begin{tabular}{lp{9.4cm}}
    \toprule   
    Predicate                           & Meaning \textit{(\code{E} event, \code{F} fluent, \code{T} timepoint)}                              \\
    \midrule                            
    \codeB{holdsAt(F,T)               } & \code{F} is true at \code{T}                                                                        \\
    \codeB{happens(E,T)               } & \code{E} occurs at \code{T}                                                                         \\
    \codeB{releasedAt(F,T)            } & \code{F} is released from inertia at \code{T}                                                       \\
    \codeB{initiates(E,F,T)           } & if \code{E} occurs at \code{T}, then \code{F} is true and not released from inertia after \code{T}  \\
    \codeB{terminates(E,F,T)          } & if \code{E} occurs at \code{T}, then \code{F} is false and not released from inertia after \code{T} \\
    \codeB{releases(E,F,T)            } & if \code{E} occurs at \code{T}, then \code{F} is released from inertia after \code{T}               \\
    \codeB{trajectory(F1,T1,F2,T2)    } & if \code{F1} is initiated by an event at \code{T1}, then \code{F2} is true at \code{T1+T2}          \\
    \codeB{antiTrajectory(F1,T1,F2,T2)} & if \code{F1} is terminated by an event at \code{T1}, then \code{F2} is true at \code{T1+T2}         \\
    \gmidrule                           
    \codeB{startedIn(T1,F,T2)         } & \code{F} is initiated by an event in the interval \code{(T1,T2)} \textit{(auxiliary pred.)}         \\
    \codeB{stoppedIn(T1,F,T2)         } & \code{F} is terminated by an event in the interval \code{(T1,T2)} \textit{(auxiliary pred.)}        \\
    \bottomrule\\[-.5em]
  \end{tabular}
\end{table}

EC specifications of problem instances consist of a~universal theory,
a~domain description, and a narrative.  The theory is a~conjunction of EC
axioms, the domain description consists of the causal laws of the
domain, 
and the narrative provides observations of event occurrences and
about fluents. 
A basic example encoded using \DEC{} follows:

\begin{example}[Counter~\linkIconDecEx{ex15-counter}~\linkIconHecEx{ex15-counter} \footnotemark{}]\label{exa:counter-base}
  Consider the example from \cite{mueller-ec-chapter} that models a counter that can be incremented or reset.
  Figure~\ref{sfig:counter-dec-domain} models a fragment (omitting the reset) of the rules of the domain, defining the fluent \code{val(V)}, the \code{inc} event, and its conditional effects on the fluent.
  Then, Figure~\ref{sfig:counter-dec-narrative} models the narrative of initial observations and event occurrences.
  For this narrative, \code{val(1)} would hold at all timepoints in the interval \code{(10,20]}.
  Finally, Figure~\ref{sfig:counter-dec-axioms} models the relevant \DEC{} axioms defining when a fluent \textit{holds}.
\end{example}

\footnotetext{We use ``\linkIconDecEx{}~\linkIconHecEx{}'' to link to our repository~\repo{} \artifactDoi{}. The left~\linkIconDecEx{} links to a traditional ASP version and the right~\linkIconHecEx{} to the Hybrid ASP version.}

\begin{figure}[tb]
\figVspaceB{} 
  \begin{subfigure}[b]{.55\textwidth}
  \begin{subfigure}[b]{.99\textwidth}
\begin{lstlisting}[style=MySCASP]
time(0..100).     dom(0..10).
fluent(val(V)) :- dom(V). 
initiates(inc,val(V+1),T) :- holdsAt(val(V),T).
terminates(inc,val(V),T)  :- holdsAt(val(V),T).
\end{lstlisting}    
        \sfigCapVspaceB{}\caption{Domain description}
        \label{sfig:counter-dec-domain}
        \vspace{-1.3em} 
  \end{subfigure}
  \begin{subfigure}[b]{.99\textwidth}
\begin{lstlisting}[style=MySCASP, firstnumber=5] 
holdsAt(val(0), 0).       % initially zero
:- holdsAt(val(V), 0), V != 0, dom(V).
happens(inc,   10).       % increment to 1
happens(inc,   20).       % increment to 2
\end{lstlisting}    
        \sfigCapVspaceB{}\caption{Narrative}
        \label{sfig:counter-dec-narrative}
  \end{subfigure}
  \end{subfigure}
  \begin{subfigure}[b]{.44\textwidth}
\begin{lstlisting}[style=MySCASP, firstnumber=9] 
holdsAt(F,T+1) :-     % event effect
  happens(E,T), initiates(E,F,T),
  time(T+1).
holdsAt(F,T+1) :-     % inertia
  holdsAt(F,T), not releasedAt(F,T+1),
  not terminated1(F,T), time(T+1).
terminated1(F,T) :-   % auxiliary
  happens(E,T), terminates(E,F,T).
releasedAt(F,T+1) :- ...
  % similar to holdsAt using releases
\end{lstlisting}    
        \sfigCapVspaceB{}\caption{Relevant \DEC{} axioms}
        \label{sfig:counter-dec-axioms}
  \end{subfigure}
\caption[]{\DEC{} encoding of the counter example.}
  \label{fig:counter-dec}
\end{figure}

\ssecVspaceB{}\subsection{Hybrid ASP}\ssecVspaceA{} 
\label{sec:hybrid-asp}

We rely on a basic acquaintance with ASP.
The syntax of our logic programs follows the one of \clingo by~\cite{potassco-user-guide};
its semantics are detailed by~\cite{abstract-gringo}.
The Hybrid ASP systems \clingcon{}
and \clingolpx{}\footnote{Available at \href{https://potassco.org/clingcon}{https://potassco.org/clingcon} and \href{https://github.com/potassco/clingo-lpx}{https://github.com/potassco/clingo-lpx}.}
extend
the input language of \clingo{} by \emph{theory atoms} representing \emph{linear constraints}
that have the form

\vspace{-1.2em} 
\begin{lstlisting}[style=MySCASP,numbers=none]
                             &sum{$\,\alpha_1$*$t_1$; ...; $\,\alpha_k$*$t_k$ } $\circ$ $\beta$
\end{lstlisting}
where $k\ge 0$, 
the $\alpha_i$'s and $\beta$ are numeral terms, 
the $t_i$'s are symbolic terms, and 
\mbox{$\circ\in\{\lstinline|<|,\lstinline|<=|,\lstinline|=|,\lstinline|>=|,\lstinline|>|,\lstinline|!=|\}$}.
Such a theory atom represents the linear constraint
\[ \alpha_1 \cdot t_1 + \ldots + \alpha_k \cdot t_k \circ \beta \]
where $t_1$, \ldots, $t_k$ serve as variables, 
and the $\alpha_i$'s and $\beta$ stand for integer constants.
For convenience,
$\alpha_i$\lstinline|*|$t_i$ may be written as $\alpha_i$ or $t_i$, 
and $\beta$ may also be of the form
$\alpha$\lstinline|*|$t$, $\alpha$ or $t$.

In \clingcon by~\cite{clingcon}, stable models are extended by
an assignment from symbolic terms to \emph{integers}. 
Theory atoms are evaluated with respect to that assignment, 
and symbolic terms are left open:
they are not required to be founded and 
can therefore take any integer value without being justified by a rule.
The semantics of \clingolpx{} are the same as for \clingcon{}, 
except that the assignment maps symbolic terms to \emph{rationals}.%
\footnote{
The current \clingolpx{} implementation 
does not allow theory atoms in the body,
but these can be simulated by theory atoms in the head and additional regular atoms.
}

These hybrid systems do not ground symbolic terms over their
possible values but instead rely on 
dedicated algorithms for linear constraints.
This reduces grounding size and often improves solving performance.
Additionally, \clingolpx{} extends the expressiveness of \clingo{} 
by supporting reasoning about rational numbers.

\begin{example}[Falling Object]\label{exa:falling-base}
  Consider the example by~\cite{mueller_book-fixed} of an object falling at constant speed. 
  The following program represents 
  the height of the object at two time steps (without EC):
  \vspace{-1.4em} 
\begin{lstlisting}[style=MySCASP]
&sum{ 0 } < time(1).
falling :- &sum{ (height,0) } > 0.
&sum{ (height,0); -2*time(1) }  = (height,1) :-     falling.
&sum{ (height,0) }              = (height,1) :- not falling.
\end{lstlisting}
  The term \code{time(1)} denotes the time passed at step \code{1}, 
  while \code{(height,0)} and \code{(height,1)} denote the height of the object at steps \code{0} 
  and \code{1}, respectively.
  The first rule requires the \code{time} at step \code{1} to be greater than \code{0}.
  The \code{height} at step \code{0} is left open. 
  If it is above \code{0}, 
  then the second rule makes \code{falling} true, 
  and consequently the third rule
  sets the \code{height} at \code{1} to be the initial \code{height}
  minus \code{2} times the time passed at step \code{1}.
  Otherwise, the fourth rule makes the \code{height} at \code{1} persist from step \code{0}.
  This program has one stable model for every value of \code{(height,0)} and 
  every value of \code{time(1)} greater than \code{0}.
  If \code{(height,0)} is above \code{0}, 
  then \code{falling} belongs to the model and 
  the value of \code{(height,1)} is given by the expression
  \code{(height,0)-2*time(1)}.
  Otherwise, 
  \code{falling} is not in the model and \code{(height,1)} has the same value as \code{(height,0)}. 
\end{example}

\secVspaceB{}\section{Related work}\secVspaceA{}
\label{sec:related-work}

Automated reasoning techniques for the Event Calculus have evolved over time. An early approach is the Prolog-based Event Calculus planner by~\cite{shanahan_abductive}.  This was later followed by the SAT-based DEC reasoner by~\cite{mueller_book-fixed}. More recently, \cite{reformulating} reformulated the Event Calculus in terms of ASP and provided the automated transformation tools EC2ASP and F2LP. This enabled the use of state-of-the-art ASP solvers, such as \clingo{} by~\cite{clingo-paper}, which offer superior performance and expressiveness compared to the earlier approaches.
All these approaches considered only discrete time and  discrete values of fluents. This restriction is inherent to SAT- and ASP-based methods, as the ground phase requires finite domains and this precludes the representation of infinite/dense domains.
Moreover, grounding can lead to the well-known grounding explosion as the size of the domain increases.

Alternative approaches for reasoning about EC, e.g., by \cite{arias22-tplp-ec} and~\cite{iclp24-pcapump-ec-scasp,iclp25-ec-scasp-zeno}, use \scasp{}, a grounding-free, query-driven ASP solver by~\cite{arias18-iclp-scasp} which supports linear constraints over rationals.  To the best of our knowledge, these are the only ones for fully-automated, general-purpose, logical reasoning about EC that reasons considering continuous change while avoiding the grounding explosion and offering multiple reasoning modes, including abduction.  However, as an academic prototype, \scasp{} lacks the optimized performance of \clingo{} and, further, can suffer from non-termination problems due to its query-driven, Prolog-style execution.

There are other systems that extend ASP with linear constraints, e.g., those based on translation-based approaches, such as EZSMT or ASPMT2SMT, and constraint propagation approaches, such as DLVHEX[CP] or EZCSP, which each have their own advantages and disadvantages---for a comprehensive survey see the work by \cite{lierler2023constraint}.

Finally, there exists a wide range of specialized applications of EC, which often use their own specialized tool for the reasoning.  Such tools typically offer much better performance but only for a particular class of problems or for specific reasoning tasks.  For example, RTEC by~\cite{rtec} supports reasoning in real-time and is specialized in monitoring streams of events to detect composite events.

\secVspaceB{}\section{Hybrid Event Calculus}\secVspaceA{}

\label{sec:from-dec-to-hec}

The main goal of our work is to reason about EC with continuous changes and, thus, with dense value domains of fluents and time.
Traditional ASP solvers, such as clingo, fully ground the program. 
On the other hand, Hybrid ASP allows one to keep only part of the model grounded, while other parts can be represented by constraints, in such a way that they are not grounded, and, thus, we can use these constraints to represent dense and/or infinite domains.
In this section, we present Hybrid EC, an extended axiomatization of \DEC{}, a formulation introduced by \cite{mueller_book-fixed} that has been proven to be equivalent to the \EC{} formulation when restricted to integers.
The core aspect of \DEC{} is that it operates \textit{timepoint by timepoint} instead of over \textit{spans of timepoints}, i.e.,
always referencing two consecutive timepoints \code{T} and \code{T+1}. 
Hybrid EC further introduces two core ideas:
\begin{enumerate}
\item The definition of \emph{functional fluents} that can hold dense values since these values (and the relations among them) can be represented by constraints and, thus, can be grounding-free (Section~\ref{sec:functional-fluents}).
\item A \textit{hybrid representation of time} where significant timepoints are mapped to discrete steps of the system and only those are grounded, while the actual time is handled by the constraint solver and is, thus, grounding-free and can be dense (Section~\ref{sec:mapping-steps}).
\end{enumerate}

In addition, we present \HEC{}, an implementation of Hybrid EC in Hybrid ASP, 
that builds on the ASP-based encoding of \DEC{} proposed by \cite{reformulating}
and utilizes the two above mentioned ideas.
The implementation incrementally executes a Hybrid ASP solver, either \clingcon{} or \clingolpx{} (Section~\ref{sec:introducing-dense-time}), gradually increasing the number of steps in the hybrid representation of time until a model is found 
(Section~\ref{sec:incremental-evaluation}).
It utilizes the Hybrid ASP encoding of our \HEC{} axioms (Section~\ref{sec:modeling-hec-axioms}) along with additional rules for handling observations defined in a narrative (Section~\ref{sec:modeling-narrative}) and events triggered via continuous change or time (Section~\ref{sec:triggered-events}).
A problem instance then needs to provide a general encoding of the domain description and a particular narrative. 

\ssecVspaceB{}\subsection{Functional Fluents}\ssecVspaceA{}
\label{sec:functional-fluents}

Fluents in EC represent time-varying properties of the world: e.g., 
\code{falling(apple)} represents that an apple is falling, and \code{height(apple,Value)} represents the height of the apple within a given range of numerical values.
Standard EC supports only relational fluents, that is, fluents are either true or false
at any given timepoint, which is represented as \code{holdsAt(Fluent,Time)} or its negation.
In \DEC{}, fluents are ground atoms, e.g., \code{height(apple,1), ..., height(apple,10)}\footnote{Each ground instance of \code{height(apple, Value)} is treated as a separate relational fluent.} 
that, in principle, can hold at the same time.
The \emph{situation calculus} by~\cite{reformulating}
supports functional fluents, which can take different values, 
e.g., \code{height(apple)=Value}.
According to \cite{mueller_book-fixed}, functional fluents can be represented by relational fluents in EC, 
e.g., \code{f(X)=Y} can be encoded as \code{f(X,Y)},
provided that for any \code{X} and \code{T} there is exactly one \code{Y} such that \code{holdsAt(f(X,Y),T)}.

We introduce functional fluents in EC using the language of Hybrid ASP.
The value of a functional fluent \code{F} at a step \code{S} is given by the symbolic term \code{(F,S)}.
A value \code{Val} can be assigned to this term using the theory atom \code!&sum{(F,S)}=Val!.
Recall that in \clingcon{} and \clingolpx{}, the values of  symbolic terms are left open. 
This corresponds closely to the relational representation of \DEC{}
by~\cite{reformulating}, 
where fluent values are left open using choice rules of the form \code!{holdsAt(F,T)}!.

\ssecVspaceB{}\subsection{Hybrid Representation of Time}\ssecVspaceA{}

\label{sec:mapping-steps}

\DEC{} uses integer time steps, and all predicates are grounded to all timepoints.
However, all changes in EC are caused by event occurrences, and
fluents either maintain a constant value or their value changes via a trajectory between occurrences.
Therefore, one can focus the reasoning only on the significant timepoints, i.e., those with event occurrences, seeing them as states.
We propose to treat \DEC{} time steps as \textit{abstract steps of the system} instead of integer increments of time.
Each step corresponds to the occurrence of an event, representing the state of the system at the time of occurrence and the state transition initiated by it.
The number of steps must be equal to the number of event occurrences at distinct timepoints plus one step for an implicit final state that allows us to observe the effects of the last event.
Each step then has a timepoint associated with it via constraints.
This means that a variable amount of time can pass between each pair of subsequent steps and that the grounding size is determined by the number of steps instead of the size of the time domain.
A limitation of this approach is that, currently, we must specify the
number of steps to be used when executing the reasoning.  For
now, assume that we guess the right number of steps, we explain an
incremental execution approach in
Section~\ref{sec:incremental-evaluation}.

\begin{figure}[tb]
\figVspaceB{} 
    \centering
    \begin{lstlisting}[style=MySCASP]
&sum{(time, 0) } >= 0.                                   % step 0 is at time 0
&sum{(time, laststep)} = maxtime.                        % last step is at maxtime
&sum{(time, S1+1) } > (time,S1) :- step(S1), step(S1+1). % ordering of consecutive steps
% T maps to S when an event is observed at T and the constraint (time,S) = T is satisfied
mapEventObs(T,S) :- obs(happens,E,T), step(S), &sum{ (time,S) } = T.
                 :- obs(happens,E,T), not mapEventObs(T,_). % all observs. must be mapped
    happens(E,S) :- obs(happens,E,T),     mapEventObs(T,S). % version of happens for steps
    \end{lstlisting}
    \caption{Mapping of (dense) time values to discrete steps.}
    \label{fig:mapping}
\end{figure}

Figure~\ref{fig:mapping} shows the encoding of this hybrid
representation, where the time at step \code{S} is denoted by 
the symbolic term \code{(time,S)}. 
Lines~1--3 
define the time of the first and last steps and ensure
that the time associated with consecutive steps 
is strictly increasing.
Within this representation, 
an atom \code{holdsAt(F,S)} expresses that a relational
fluent \code{F} holds at step~\code{S}.
For \code{S=0},
this means that \code{F} holds 
in the time interval \code![0,(time,0)]!,
while for \code{S>0}, 
\code{F} holds 
in
\code!((time,S-1),(time,S)]!.
Note that \code{(time,0)} may be greater than zero. 
Fluents that undergo continuous change have a specific value at each step \code{S}, but their value changes between steps 
according to the corresponding trajectory.

Figure~\ref{fig:mapping} also shows 
how observations of event occurrences, which come as facts in the narrative of a problem instance, are mapped to time steps.
An observation of an event~\code{E} 
at a timepoint \code{T} is represented by an atom 
\code{obs(happens,E,T)}
(see Section~\ref{sec:modeling-narrative} for more details on observations).
Lines~5--6 map timepoint \code{T} to some step \code{S}, 
and line~7 derives the atom \code{happens(E,S)},
which is used by the axioms introduced in the next section.

\ssecVspaceB{}\subsection{Modeling the Axioms of Hybrid EC}\ssecVspaceA{}
\label{sec:modeling-hec-axioms}

This section explains how we extend the \DEC{} axioms as implemented by \cite{reformulating} to Hybrid EC. The extension builds on the previously defined functional fluents and the hybrid representation of time as discrete steps. This proposal provides a formal foundation for representing dense domains of fluents and time within a constraint-based framework, enabling a systematic encoding of axioms that support reasoning under Hybrid EC.
The resulting axioms, referred to as \HEC[~N] (where \texttt{N} denotes the id of the corresponding \DEC{} axiom), are summarized below:\footnote{\label{foot:axioms}We include the original \DEC{}~\linkIconDecAx{} axioms in Appendix~\ref{apx:sec:dec-axioms} and the full \HEC{}~\linkIconHecAx{} axioms in Appendix~\ref{apx:sec:hec-axioms}.} 

\begin{description}
\item[\textbf{\HEC[~1]}] redefines \code{stoppedIn} as in \DEC{} only for relational fluents.
\item[\textbf{\HEC[~2]}] extends \code{startedIn} for both types of fluents.
\item [\textbf{\HEC[~3-4]}] capture the transition from gradual change of relational fluents to continuous change of functional fluents via \code{trajectory}/\code{antiTrajectory}.
\item [\textbf{\HEC[~5]}] extends positive inertia for values of functional fluents and \code{holdsAt} of relational fluents.
\item [\textbf{\HEC[~6]}] redefines negative inertia of \code{holdsAt} as in \DEC only for relational fluents.
\item [\textbf{\HEC[~7-8]}] redefines (the absence of) \code{releasedAt} capturing when any type of fluent is released or ``un-released'' from inertia by an event.  
\item [\textbf{\HEC[~9]}] extends positive effects of events via \code{initiates} for both types of fluents.
\item [\textbf{\HEC[~10]}] redefines negative effects of events via \code{terminates} only for relational fluents.
\item [\textbf{\HEC[~11-12]}] extends inertia for (the absence of) \code{releasedAt} for both types of fluents.
\end{description}

For brevity, we only focus on the implementation details of two most important sets of axioms: (1) In Section~\ref{ssec:modeling-hec-axioms:effects-and-inertia}, we explain how to capture the effects of events and inertia on functional fluents. (2) In Section~\ref{ssec:modeling-hec-axioms:trajectory}, we explain the transition from gradual change of relational fluents to continuous change of functional fluents.

\sssecVspaceB{}\subsubsection{Axioms for Event Effects and Inertia}\sssecVspaceA{}
\label{ssec:modeling-hec-axioms:effects-and-inertia}
\begin{figure}[tb]
\figVspaceB{} 
\begin{lstlisting}[style=MySCASP]
%% HEC 5: transformed DEC 5 to steps and added a clause for functional fluents
holdsAt(F,S+1) :-  holdsAt(F,S), not terminated1(F,S), not releasedAt(F,S+1), step(S+1).  
&sum{(F,S)} = (F,S+1) :-
  ffluent(F), step(S), not initiated1(F,S), not releasedAt(F,S+1), step(S+1).
%% HEC 9: transformed DEC 9 to steps and added a clause for functional fluents
holdsAt(F,S+1) :- happens(E,S), initiates(E,F,S), step(S+1).
&sum{C*V : @member(LE) = (C,V)} = (F,S+1) :- happens(E,S), initiates(E,F,LE,S), step(S+1).
%% Auxiliary
terminated1(F,S) :- happens(E,S), terminates(E,F,S).
initiated1(F,S) :- happens(E,S), initiates(E,F,_,S).                     % extra arg. '_'
    \end{lstlisting}
  \caption[]{Axioms of \HEC[\,5] and \HEC[\,9] representing event effects and inertia.}
  \label{fig:axioms-effect}
\end{figure}
\noindent
In \DEC{}, a relational fluent starts to hold after being \textit{initiated} by an event, ceases to hold when
\textit{terminated} by an event, or persists by \textit{inertia} unless it is \textit{released}.
This behavior is encoded by~\cite{reformulating} using axioms \DEC[\,5-6] for inertia and axioms \DEC[\,9-10] for event effects (\DEC{\,5} and \DEC{\,9} were shown in Figure~\ref{sfig:counter-dec-axioms}).
In \HEC{}, these axioms must be adapted to handle both relational and functional fluents (see Figure~\ref{fig:axioms-effect}).
For relational fluents, this adaptation is straightforward, as we simply replace predicate \code{time} by \code{step} (lines 2 and 6 of~Figure~\ref{fig:axioms-effect}).

For functional fluents, the rule in lines~3-4 of \HEC[\,5] represents inertia. 
The theory atom in the head assigns to fluent \code{F} at step \code{S+1} the value of \code{F} at step \code{S} 
unless the fluent is initiated or released from inertia.
Line~7 of \HEC[\,9] represents the effect of initiating fluent \code{F}
whenever event \code{E} happens at step \code{S}.
The corresponding theory atom assigns to fluent \code{F} at step \code{S+1}
the value of the linear expression \code{LE}, given by a new predicate \code{initiates/4}.
%
A~linear expression
\code{c1*v1+...+cN*vN}
is encoded as a tuple
\code{((c1,v1),...,(cN,vN))},\footnote{A tuple with a single element is encoded with a comma at the end, e.g. \code{((c1,v1),)}.}
which is unpacked by the Python external function \code{@member},
yielding the set of theory elements
\code!{(c1*v1);...;(cN*vN)}!.
Rules for \HEC[\,6] or \HEC[\,10] about the termination of fluents are not needed,
since functional fluents have exactly one value at every time step, 
and initiating a new value breaks the inertia of the previous value by deriving \code{initiated1} (line~10).

\begin{example}[Counter, Example~\ref{exa:counter-base} cont.~\linkIconDecEx{ex15-counter}~\linkIconHecEx{ex15-counter}]\label{exa:counter-hec}
To demonstrate how to encode the effects of events on functional fluents we transform the \DEC encoding of the domain description of the counter example from Figure~\ref{fig:counter-dec} into \HEC as follows (see the full encoding in Appendix~\ref{apx:example:counter}):
\vspace{-1.4em} 
\begin{lstlisting}[style=MySCASP]
ffluent(val).                    % argument V moved into constraints and no domain needed
initiates(inc,val,((1,(val,S)),(1,1)),S) :- step(S).    % LE: (val,S+1) = 1*(val,S) + 1*1    
\end{lstlisting}    
\vspace{-0.5em} 
\end{example}

\sssecVspaceB{}\subsubsection{Axioms for Continuous Change}\sssecVspaceA{}
\label{ssec:modeling-hec-axioms:trajectory}

Fluents in EC can undergo continuous change via trajectories and anti-trajectories.
\DEC{} can only model gradual change over discrete steps. 
In contrast, Hybrid EC supports reasoning about dense domains, and 
can faithfully model continuous change, although it is limited to monotonic linear changes.

Figure~\ref{sfig:axioms-traject-dec} shows axiom \DEC[\,3] for gradual change in \DEC{} using trajectories.
Figure~\ref{sfig:axioms-traject-hec} shows the corresponding rule \HEC[\,3] for continuous change, which only applies to functional fluents as there are no trajectories for relational fluents in \HEC. 
The representation of anti-trajectories in \DEC[\,4] and \HEC[\,4] is analogous and we do not discuss it further. 
In \HEC{\,3}, the value of a functional fluent~\code{F2} at step \code{S1+S2} is defined by a new predicate \code{trajectory(F1,S1,F2,(LE,R),S2)}
if an event~\code{E} initiated the relational fluent~\code{F1} at step~\code{S1} and it has not been stopped since.
The new atom over the predicate~\code{trajectory} 
specifies the value of \code{F2} using the tuple \code{(LE,R)}.
The linear expression \code{LE} provides the value of \code{F2} at the beginning of the trajectory, 
and \code{R} gives the changing rate during the trajectory.\footnote{\code{R} is separated from \code{LE} so that it can be used in the head of \HEC[\,3] and for observations in Section~\ref{sec:modeling-narrative}.} 
The new atom further refers to step \code{S1} and \textit{duration in steps} \code{S2} instead of timepoints.
This is a key difference because the value assigned by the trajectory to~\code{F2} cannot be derived directly from the step-based duration, as the result would vary if additional steps are inserted (e.g., due to an unrelated event occurring during the trajectory).
Therefore, the theory atom in the head
represents the linear expression \code{LE} as in \HEC[\,9] (unpacked by the function \code{@member}) but further
adds
\code{R*(time,S1+S2)} and \code{-R*(time,S1)}
to encode the change
\code{R*((time,S1+S2)-(time,S1))}, which refers to the actual time difference between the concerned steps.

\begin{figure}[tb]
\figVspaceB{} 
  \begin{subfigure}[t]{.47\textwidth}
    \vspace{1.2em} 
    \begin{lstlisting}[style=MySCASP]
holdsAt(F2,T1+T2) :- 
 trajectory(F1,T1,F2,T2),
 0<T2, happens(E,T1), initiates(E,F1,T1),
 not stoppedIn(T1,F1,T1+T2), time(T1+T2).
\end{lstlisting}
    \sfigCapVspaceB{}\caption{Original axiom \DEC[\,3]}
    \label{sfig:axioms-traject-dec}
  \end{subfigure}
  \begin{subfigure}[t]{.51\textwidth}
    \begin{lstlisting}[style=MySCASP]
&sum{C*V: @member(LE)=(C,V); R*(time,S1+S2); 
      -R*(time,S1)} = (F2,S1+S2) :- 
 trajectory(F1,S1,F2,(LE,R),S2),
 0<S2, happens(E,S1), initiates(E,F1,S1),
 not stoppedIn(S1,F1,S1+S2), step(S1+S2).
    \end{lstlisting}
    \sfigCapVspaceB{}\caption{Replacement axiom \HEC[\,3]}
    \label{sfig:axioms-traject-hec}
  \end{subfigure}
  \vspace{0.2em} 
  \caption[]{Axioms for representing gradual (\DEC) and continuous (\HEC) change.}
  \label{fig:axioms-traject}
\end{figure}

\begin{example}[Falling Object, Example~\ref{exa:falling-base} cont.~\linkIconDecEx{ex3-falling_object}~\linkIconHecEx{ex3-falling_object}]\label{exa:falling-trajectory}
To show the transition from gradual to continuous change, we encode the trajectory of the falling object both in \DEC and in \HEC as follows: (see the full encoding in Appendix~\ref{apx:example-falling-object}) 
\vspace{-1.4em} 
\begin{lstlisting}[style=MySCASP, basewidth=.43em]
trajectory(falling(O),T1,height(O,H2),T2) :- holdsAt(height(O,H),T1), H2 = H-2*T2, %% DEC
  time(T2), time(T1+T2).
trajectory(falling(O),S1,height(O),(((1,(height(O),S1)),),-2),S2) :-               %% HEC
  object(O), step(S1), step(S2), step(S1+S2).
\end{lstlisting}
The trajectory starts when the fluent \code{falling(O)} is initiated by a \code{drop(O)} event, which also releases \code{height(O)} from inertia, and ends when \code{falling(O)} is terminated.
\end{example}

\ssecVspaceB{}\subsection{Modeling Narratives in Hybrid EC}\ssecVspaceA{}
\label{sec:modeling-narrative}

A narrative of a problem instance consists of observations 
of event occurrences and fluent values.
In \HEC{}, observations are represented by facts of the form 
\code{obs(O1,O2,T)} where~\code{T} denotes a timepoint:
\code{obs(happens,E,T)} for event occurrences,
\code{obs(holdsAt,F,T)} and \code{obs(notHoldsAt,F,T)} for values of relational fluents,
\code{obs(F,Val,T)} for values of functional fluents, and
\code{obs(releasedAt,F,T)} and \code{obs(notReleasedAt,F,T)} 
for fluents released from inertia.
We have already discussed event observations in Section~\ref{sec:mapping-steps}.
In this section, we focus on the representation of the other (fluent) observations.

The main task is to map observation timepoints to the steps used in \HEC.
Recall that fluents do not change between steps unless they are under the effect of a trajectory. 
As discussed in Section~\ref{sec:mapping-steps}, 
the value of a fluent at step \code{S} 
represents its value over the time interval \code{((time,S-1),(time,S)]}.
Hence, any observation at a timepoint within \code{((time,S-1),(time,S)]} can be checked at step \code{S}.
Based on this, the encoding maps timepoints to their corresponding steps
and checks the observations on those steps.
The mapping is defined in lines 2--5 of Figure~\ref{fig:observations} using predicate \code{mapObs(T,S)}.
The first rule covers the initial step, 
while the second rules handles the case where \code{T} falls within \code{((time,S-1),(time,S)]} for \code{S > 0}.
The constraint in line~7 uses this predicate to encode 
the observation of a relational fluent \code{F} 
holding at timepoint \code{T}.
It forbids cases where \code{T} is mapped to step \code{S} but \code{F} does not hold at \code{S}.
Other observations of relational fluents are encoded analogously.\footnote{Alternatively, we could introduce new steps for observations, 
as for event occurrences.
However, since the grounding size of \HEC{} depends on the number of steps, 
we chose not to introduce additional steps.}

For functional fluents,
the auxiliary predicate \code{in_ttory(F,S,R)}, defined in lines 12-13, 
identifies the case where fluent \code{F} is under a trajectory at step \code{S}, 
and extracts the rate of change \code{R}. 
If no trajectory applies to \code{F} at step \code{S},
the situation is analogous to relational fluents, and
line 9 checks the observed value of \code{F} at the 
corresponding step \code{S}.
Otherwise, 
line 10 computes the value of \code{F} at the time \code{T} of the observation
as \code{(F,S)-R*((time,S)-T))}, 
and checks it against the observed value \code{V}.

\begin{example}[Falling Object, Example \ref{exa:falling-trajectory} cont.~\linkIconDecEx{ex3-falling_object}~\linkIconHecEx{ex3-falling_object}]\label{exa:falling-observations}
  Assume step \code{0} is mapped to time \code{10},
  step \code{1} is mapped to time \code{20},
  object \code{obj} has height \code{20} at step \code{0}, 
  and height \code{0} at step \code{1}.
  That is, \code{(time,0)=10}, \code{(time,1)=20}, 
  \code{(height(obj),0)=20}, and \code{(height(obj),1)=0}.
  Observation \code{obs(holdsAt,falling(obj),15)} holds 
  because timepoint \code{15} is mapped to step \code{1}, and 
  \code{falling(obj)} is true at that step. 
  Indeed, \code{obj} is always falling during the time between steps \code{0} and \code{1}.
  Observation \code{obs(height(obj),10,15)} also holds. 
  The height at step \code{1} is \code{0}, 
  but the height at time \code{15} is \code{20-2*(20-15)=10} due to the trajectory,
  which is consistent with the observation.
\end{example}

\begin{figure}[tb]
\figVspaceB{} 
    \centering
    \begin{lstlisting}[style=MySCASP]
% observation at T applies to a step with the closest larger time
mapObs(T,S) :- obs(O,_,T), O != happens, S = 0, &sum{( (1,(time,S )) )} >= T.
mapObs(T,S) :- obs(O,_,T), O != happens, step(S), S > 0, 
                  &sum{( (1,(time,S-1)) )} < T, &sum{( (1,(time,S )) )} >= T.
:- obs(O,_,T), O != happens, not mapObs(T, _). % every non-event must have a step
% Relational fluents' observations
:- obs(holdsAt,F,T), fluent(F), mapObs(T,S), not holdsAt(F,S). % ... other obs similar ...
% Functional fluents' observations
&sum{(F,S)} = V :- obs(F,V,T), ffluent(F), mapObs(T,S), not in_ttory(F,S,_).
&sum{(F,S); -R*(time,S); R*T} = V :- obs(F,V,T), ffluent(F), mapObs(T,S), in_ttory(F,S,R).
% F2 is in a trajectory at step S1+S2 with rate R
in_ttory(F2,S1+S2,R) :- happens(E,S1), initiates(E,F1,S1), 0<S2,
  trajectory(F1,S1,F2,(LE,R),S2), not stoppedIn(S1,F1,S1+S2), step(S1+S2).
    \end{lstlisting}
    \caption{Non-event observations.}
    \label{fig:observations}
\end{figure}

\ssecVspaceB{}\subsection{Modeling Triggered Events in Hybrid EC}\ssecVspaceA{}

\label{sec:triggered-events}
Special considerations must be made for \emph{triggered events}, i.e., events whose occurrences are not given as observations in the narrative but rather implied by the rules of the domain description. 
For such events, there will be no mapping from observations to steps (Section~\ref{fig:mapping}). 
We handle triggered events by giving the reasoning extra \textit{free-floating steps} that will not be mapped to an event occurrence observation.
The event is triggered at one of the free-floating steps which is then assigned a time according to the evaluation of linear constraints defined in the trigger rule.
However, additional constraints are needed in order to ensure that the step will not be placed at a different timepoint or that the step is not taken up by another triggered event that occurs later.

The above is a common problem in hybrid systems 
and can be addressed using a constraint that prevents steps from jumping over particular values of interest when dealing with monotonic changes, as was discussed by~\cite{shin-davis-continuous-change-in-sat} for SAT-based planning.
In our case, this is relevant for events triggered based on continuous change of fluents 
or triggered after some delay. 
We define a predicate \code{no_jump(Fluent, Step, Value)} that introduces such a constraint for typical scenarios, 
as shown in Figure~\ref{fig:no_jump} (lines~1--2).
The body of the rule that implies \code{no_jump} should mirror the body of the trigger rule.

The implicit last step (discussed in Section~\ref{sec:mapping-steps}) is crucial to force the reasoning to check that the \code{no_jump} constraints are not violated after the last event occurrence (second to last step), which would signal the need for more steps.

\begin{example}[Falling Object, Example \ref{exa:falling-observations} cont.~\linkIconDecEx{ex3-falling_object}~\linkIconHecEx{ex3-falling_object}]\label{exa:falling-trigger}
Consider, the falling object that is dropped from an initial height of 20 at time 10 and hits the ground when its height reaches zero.
The observation of dropping the object and the triggered event of hitting the ground are encoded on lines~4--5 of Figure~\ref{fig:no_jump}.
This problem would need 3 steps in \HEC: drop, \textit{hit the ground}, and the last step.
Reaching height zero is defined by the falling trajectory which in this case implies a trigger time of~20.
Line 6 shows the necessary \code{no_jump} constraint saying that there must not exist two consecutive steps such that the height is less than zero at the first one and greater than zero at the second one, i.e., one of the steps must have height exactly zero.
Note that the object must be falling at the second step and not necessarily at the first step, as the value of a fluent at step \code{S} represents the time period \code{((time,S-1),(time, S)]}.
\end{example}

\begin{figure}[tb]
\figVspaceB{} 
    \centering
    \begin{lstlisting}[style=MySCASP]
:- no_jump(F,S,V), &sum{ (F,S-1) } < V, &sum{ (F,S) } > V. % cant step over V going up
:- no_jump(F,S,V), &sum{ (F,S-1) } > V, &sum{ (F,S) } < V. % cant step over V going down

obs(happens(drop(obj)), 10). 
happens(hitGround(O), S) :- holdsAt(falling(O), S), &sum{ (height(O), S) } = 0.
no_jump(height(O), S, 0) :- holdsAt(falling(O), S). % no jump over height 0 while falling
    \end{lstlisting}
    \caption{The \code{no_jump} constraint and a triggered event for the falling object.}
    \label{fig:no_jump}
\end{figure}

\ssecVspaceB{}\subsection{Incremental Solving}\ssecVspaceA{}
\label{sec:incremental-evaluation}

Mapping time to steps (Section~\ref{sec:mapping-steps}) requires at least as many steps as there are observations of event occurrences at distinct timepoints. However, it might be hard to know how many triggered events will occur in a particular problem instance.
We have designed the axioms in such a way that the reasoning will only produce models for the right number of steps to cover all the events that occur from time zero to the time of the implicit last step, i.e., there will be no models for not enough steps and for too many steps.

This is achieved by a combination of three techniques: (1) the implicit last step, (2) the \code{no_jump} constraints, and (3) a significant step constraint.
The \code{no_jump} constraints and the implicit last step (Section~\ref{sec:mapping-steps}) ensure that no models exist when too few steps are provided.
To avoid excessive floating steps when too many steps are given, we introduce the constraint encoded in Figure~\ref{fig:significant_step} requiring all steps to be significant. A step is considered \emph{significant} if it involves an event occurrence or if it is the last step.
\begin{figure}[tb]
\figVspaceB{} 
    \centering
    \begin{lstlisting}[style=MySCASP]
:- step(S), not significant_step(S).   % all steps must be significant
significant_step(S) :- S=laststep.     % last step is significant
significant_step(S) :- happens(E,S).   % step is significant if an event happens
    \end{lstlisting}
    \caption{The significant step constraint.}
    \label{fig:significant_step}
\end{figure}

Our implementation uses an incremental execution approach to find the right number of steps automatically.
A limitation of this approach is that it currently is not possible to automatically identify that there is no model for any number of steps. 
Instead, the script currently explores up to a given maximum number of steps as defined by the user.
\secVspaceB{}\section{Accurately Representing Continuous Change}\secVspaceA{}
\label{sec:introducing-dense-time}

The use of integer
time and values of fluents in \DEC{} means that it cannot model continuous change. Instead, it models gradual change which can lead to inaccurate representation.

\begin{example}[Falling Object, Example \ref{exa:falling-trigger} cont.~\linkIconDecEx{ex3-falling_object}~\linkIconHecEx{ex3-falling_object}]
Consider the falling object. If it is dropped from height~20 at time~10 and falls via a constant speed of~2~units, then it will 
hit the ground at time~20 with height 0.
However, if it was dropped from height~21, then its height would be~1 at time~20 and~-1 at time~21, i.e., there would be no integer time at which the height is equal to zero.
\end{example}

This problem can be mitigated on a case-by-case basis by approximation (e.g., triggering before or after stepping over height zero), which leads to inaccurate representation, or by increasing the precision of the discretization (e.g., multiplying the time and height by a factor of 10), which can quickly lead to very large value domains. 
Although a large value domain is no longer a problem for \HEC{} with \clingcon{}, it can be difficult or impossible to identify how fine the discretization needs to be for some problem instances.

Since \HEC{} reasons about values of functional fluents and time using theory atoms, we can switch to reasoning using dense values instead of integers, which would be impossible without the hybrid approach.
We can do this by simply replacing \clingcon{} with \clingolpx{}, which reasons about theory atoms in the domain of rational numbers (see Section~\ref{sec:hybrid-asp}).

However, one must consider the challenges that come with reasoning in dense time, such as \emph{Zeno behavior} and various \emph{Zeno-like phenomena}, whose causes in EC were studied by \cite{iclp25-ec-scasp-zeno}.
For example, preventing the \emph{repeated triggering of events}, a basic modeling technique for ensuring that an event is only triggered once after its trigger condition starts to hold, becomes a fundamental problem because one may end up reasoning about the smallest timepoint \code{T2} such that \code{T2 > T1} for some given timepoint~\code{T1}, with no such \code{T2} existing in a dense domain.
These challenges can be addressed using the techniques proposed by~\cite{iclp25-ec-scasp-zeno}.
Although a detailed discussion of this in the context of \HEC{} is outside the scope of this paper and left for future work, we provide encodings of four such problems (listed in Appendix~\ref{apx:sec:comp-expr-dec}).

\secVspaceB{}\section{Experimental Evaluation}\secVspaceA{}
\label{sec:validation}

The key features to evaluate for Hybrid EC are: (1) its expressive power to encode and reason about  continuous change in EC, and (2) improvements in time and memory scalability, particularly, with respect to the size of the time and fluent value domains.

\ssecVspaceB{}\subsection{Expressiveness}\ssecVspaceA{}
\label{sec:validation-expressiveness}

We used \HEC to encode benchmark problems from the works by~\cite{reformulating} and~\cite{mueller-ec-through-sat}, additional problems from the works by~\cite{mueller-ec-chapter,mueller_book-fixed}, relevant problems from the work by~\cite{iclp25-ec-scasp-zeno}, and some modifications to highlight the benefits of using \HEC{} under rationals.
The implementations for \clingo{}, \clingcon{}, and \clingolpx{} of over 30 problems are available at \repo{} and archived at \artifactDoi{}; 
for brevity,
Table~\ref{fig:tb-comparison-expressiveness} lists only a subset 
(the full table is available in Appendix~\ref{apx:sec:comp-expr-dec}).
The problems considered illustrate the modeling and reasoning capabilities of the compared approaches.
They are grouped by their most notable features, including the reasoning mode they use, with two examples showcasing abductive reasoning.
Classification of features and reasoning follows the one used by \cite{mueller-ec-through-sat, mueller_book-fixed}.
The table confirms that \HEC{} with both \clingcon{} and \clingolpx{} can reason about all the examples that \DEC{} with \clingo{} handles. 
In addition, the use of rationals in \clingolpx{} allows us to accurately reason about problems that require continuous change, i.e., dense domains for time and fluent values, which is not possible with \clingo{} and neither with \clingcon{}.

\newcommand{\yes}{{\color{blue} \normalsize \checkmark}}
\newcommand{\yesbut}{{\color{orange} \normalsize \checkmark}}
\newcommand{\no}{{\color{red} \normalsize \texttimes}}
\begin{table}[tb]
  \setlength{\tabcolsep}{4pt}
  \linespread{0.9}
  \centering
  \smaller
  \caption{Comparing the expressiveness of \DEC{} and \HEC{}.} 
  \label{fig:tb-comparison-expressiveness}
  \begin{tabular}{p{5.2cm}lccccc}
    \toprule                                                                                                                                                                      
    \multirow{2}{*}{Reasoning Type and Notable Features} & \multirow{2}{*}{Problem}                                                                                                           & \DEC{}~\linkIconDecAx{}     & \multicolumn{2}{c}{\HEC{}~\linkIconHecAx{}} \\
                                                         &                                                                                                                                    &     \clingo{} &      \clingcon{} &      \clingolpx{} \\
    \midrule                                             
    D: state constraint                                  & \linkIconDecExECASP{DeadOrAlive40-ea.lp}~\linkIconHecExECASP{DeadOrAlive40-ea.lp}                         DeadOrAlive              & \yes       & \yes          & \yes            \\  
                                                         & \linkIconDecExECASP{Happy40-ea.lp}~\linkIconHecExECASP{Happy40-ea.lp}                                     Happy                    & \yes       & \yes          & \yes            \\  
                                                         & \linkIconDecExECASP{StuffyRoom40-ea.lp}~\linkIconHecExECASP{StuffyRoom40-ea.lp}                           StuffyRoom               & \yes       & \yes          & \yes            \\  
    \gmidrule                                            
    D: conditional effect axiom                          & \linkIconDecExECASP{Telephone40-ea.lp}~\linkIconHecExECASP{Telephone40-ea.lp}                             Telephone                & \yes       & \yes          & \yes            \\  
                                                         & \linkIconDecExECASP{Yale40-ea.lp}~\linkIconHecExECASP{Yale40-ea.lp}                                       Yale                     & \yes       & \yes          & \yes            \\  
    \gmidrule                                            
    D: effect constraint                                 & \linkIconDecExECASP{WalkingTurkey40-ea.lp}~\linkIconHecExECASP{WalkingTurkey40-ea.lp}                     WalkingTurkey            & \yes       & \yes          & \yes            \\  
    \gmidrule                                            
    \multirow[t]{3}{5.2cm}{M: non-determinism            
    (release from inertia / determining fluent)}         & \linkIconDecExECASP{RussianTurkey40-ea.lp}~\linkIconHecExECASP{RussianTurkey40-ea.lp}                     RussianTurkey            & \yes       & \yes          & \yes            \\  
                                                         & \linkIconDecExECASP{ChessBoard40-ea.lp}~\linkIconHecExECASP{ChessBoard40-ea.lp}                           ChessBoard               & \yes       & \yes          & \yes            \\  
                                                         & \linkIconDecExECASP{CoinToss40-ea.lp}~\linkIconHecExECASP{CoinToss40-ea.lp}                               CoinToss                 & \yes       & \yes          & \yes            \\  
    \gmidrule                                            
    D: concurrent event                                  & \linkIconDecExECASP{Supermarket40-ea.lp}~\linkIconHecExECASP{Supermarket40-ea.lp}                         SuperMarket              & \yes       & \yes          & \yes            \\  
    \gmidrule                                            
    A: abducible event occurrence                        & \linkIconDecExECASP{StolenCar40-ea.lp}~\linkIconHecExECASP{StolenCar40-ea.lp}                             StolenCar                & \yes       & \yes          & \yes            \\  
    \gmidrule                                            
    A: abd. fluent, disjunctive event trigger            & \linkIconDecExECASP{BusRide40-ea.lp}~\linkIconHecExECASP{BusRide40-ea.lp}                                 BusRide                  & \yes       & \yes          & \yes            \\  
    \gmidrule                                            
    D: compound event                                    & \linkIconDecExECASP{Commuter15-ea.lp}~\linkIconHecExECASP{Commuter15-ea.lp}                               Commuter                 & \yes       & \yes          & \yes            \\  
    \gmidrule                                            
    D: causal constraint                                 & \linkIconDecExECASP{ThielscherCircuit40-ea.lp}~\linkIconHecExECASP{ThielscherCircuit40-ea.lp}             ThielscherCircuit        & \yes       & \yes          & \yes            \\  
    \gmidrule                                            
    \multirow[t]{3}{4.35cm}{D: gradual change (integer), 
    event trigger, release from inertia}                 & \linkIconDecExECASP{FallingObjectWithEvents40-ea.lp}~\linkIconHecExECASP{FallingObjectWithEvents40-ea.lp} FallingObject            & \yes       & \yes          & \yes            \\  
                                                         & \linkIconDecExECASP{KitchenSink_M40-ea.lp}~\linkIconHecExECASP{KitchenSink_M40-ea.lp}                     KitchenSink              & \yes       & \yes          & \yes            \\  
                                                         & \linkIconDecExECASP{HotAirBalloon40-ea.lp}~\linkIconHecExECASP{HotAirBalloon40-ea.lp}                     HotAirBalloon            & \yes       & \yes          & \yes            \\  
    \gmidrule                                            
    \multirow[t]{3}{4.5cm}{D: continuous change          
    (rational), event trigger, release from inertia}     & \linkIconDecEx{ex3-falling_object}~\linkIconHecEx{ex3-falling_object}                                     FallingObject            & \no        & \no           & \yes            \\  %
                                                         & \linkIconDecExECASP{KitchenSink_M40-ea.lp}~\linkIconHecExECASP{KitchenSink_M40-ea.lp}                     KitchenSink              & \no        & \no           & \yes            \\  
                                                         & \linkIconDecExECASP{HotAirBalloon40-ea.lp}~\linkIconHecExECASP{HotAirBalloon40-ea.lp}                     HotAirBalloon            & \no        & \no           & \yes            \\  
    \bottomrule\\[-.5em]
\multicolumn{5}{c}{\small  Reasoning types: D=deduction, M=model finding, A=abduction.}
  \end{tabular}
\end{table}

\ssecVspaceB{}\subsection{Performance}\ssecVspaceA{}
\label{sec:validation-performance}
We now present two experiments performed on an AMD EPYC 9124 CPU (base clock of 3 GHz), measuring execution times and memory usage averaged over 10 runs with a timeout of 300 seconds while finding all models.\footnote{For the StolenCar problem, we search up to 10 models because abducing events leads to many models.}

\begin{figure}[tb]
  \begin{subfigure}[b]{.49\textwidth}
   \includegraphics[width=\textwidth]{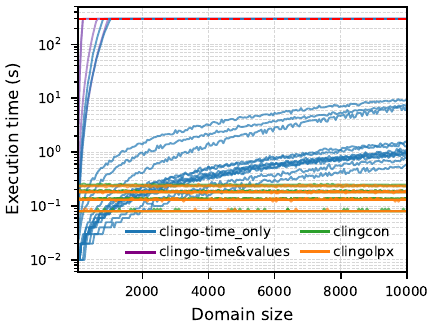}
    \caption{Scaling with the size of the domain.}
    \label{fig:comparison-execution-domain}
  \end{subfigure}
  \begin{subfigure}[b]{.49\textwidth}
    \includegraphics[width=\textwidth]{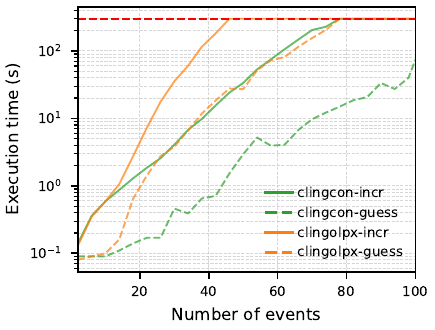}
    \caption{Scaling with the number of events.}
    \label{fig:comparison-execution-falling}
  \end{subfigure}

\end{figure}

\parVspaceB{}\paragraph{a) Domain Size:} Figure~\ref{fig:comparison-execution-domain}
shows the execution time scaling for \clingo{}, \clingcon{}, and \clingolpx{} with the domain size (time and values of fluents) using problems from the work by~\cite{reformulating} and Example~\ref{exa:counter-base} the counter (full encoding is available in Appendix~\ref{apx:example:counter}).
For plain ASP (clingo), we highlight problems that feature fluents with numerical values using purple lines since for these, the domain size increases not only in time but also in values of the fluents. The two blue lines that reach a timeout correspond to the StolenCar 
and the ThielscherCircuit 
(detailed Figures are available in Appendix~\ref{apx:sec:full-scaling-clingo}).
The results show that the domain size indeed does not affect the Hybrid ASP execution at all, while plain ASP does not scale.
This is a great advantage of using the hybrid approach, since plain ASP is fundamentally not equipped to handle large domains which greatly limits its usability in reasoning about requirements specifications of CPSs.
The memory usage scaling looks very similar to the execution time scaling with the hybrid approach using up to 30\,MB, and plain ASP using up to 50\,GB for the curves that reach the timeout and up to 1\,GB for the others (see Figure~\ref{apx:fig:comparison-execution-domain-mem} of Appendix~\ref{apx:sec:memory-scaling}).

\captionsetup[subfigure]{labelformat=mysubfig}
\captionsetup[subfigure]{labelformat=parens}

\parVspaceB{}\paragraph{b) Number of Events:} To measure the impact of including more events, we use Example~\ref{exa:falling-base} -- the falling object (full encoding is available in Appendix~\ref{apx:example-falling-object}), increasing the number of times the object is dropped.
Figure~\ref{fig:comparison-execution-falling} shows the execution time scaling
with the number of events, which is particularly relevant for evaluating the scalability of Hybrid ASP solvers, as an  increasing number of events raises both the grounding size (number of steps in \HEC) and the complexity of the reasoning.
For \clingcon{} and \clingolpx{}, we report (i)~the cost of an incremental evaluation to determine  the correct number of steps to model the problem and (ii)~a~single evaluation using the correct (guessed) number of steps.
The resulting measurements show that \clingolpx{} is slower than \clingcon{} but overall scales similarly, meaning that the introduction of rationals does not significantly impact scaling. 
In addition, the incremental execution approach shows scaling similar to that of guessing the right number of steps.
The memory usage scaling again looks similar to the execution time scaling with the maximum reaching around 300\,MB.
A notable difference is that \clingolpx{} only uses more memory than \clingcon{} for up to around 30\,events and uses less from then on (see Figure~\ref{apx:fig:comparison-execution-falling-mem} of Appendix~\ref{apx:sec:memory-scaling}).

Clearly, unlike the domain size, the number of events does have a significant impact on the hybrid approach, which is complementary to the results of \clingo{} (not shown in the graph not to clutter it too much), on which the number of events has almost no impact since the  domain size remains unchanged.
However, we note that, in the context of high-level requirements specifications of CPSs, it is often much more important to be able to handle large domains and accurately represent continuous behavior rather than to represent large numbers of events. 
Indeed, many errors can be detected within relatively short narratives corresponding to use cases described as a part of the specification as shown, e.g., in the work by~\cite{iclp24-pcapump-ec-scasp}, while restricting the domain or discretization can lead to missing the error.
Finally, we note that, compared to \scasp{}, the hybrid approach provides a significant speedup since \scasp{} reaches the timeout on the problem considered in Figure~\ref{fig:comparison-execution-falling} at already eight events.

\secVspaceB{}\section{Conclusion}\secVspaceA{}
\label{sec:conclusion}

We have presented Hybrid EC, an encoding of the Event Calculus for Hybrid ASP solvers based on \DEC{}, and its implementation in \HEC{} axioms.
The resulting framework combines ASP with linear constraints (over integers and/or rationals) that effectively avoids the grounding bottleneck when the domain of time and fluent values increases, as well as the problem of inaccurate modeling due to discretization when reasoning considering continuous changes (requiring dense domains).  We have shown both of these advantages on a number of problems from the literature 
and provide all their encodings as modeling guides at \repo{} \artifactDoi{}.
    
This work opens several lines for future research, e.g.: using \HEC{} to analyze real-life CPS specifications as done with \scasp{} for the PCA pump specification in the work by~\cite{iclp24-pcapump-ec-scasp}, hopefully avoiding some of the scalability/termination barriers hit with \scasp{}; a deeper comparison of \HEC{} for \clingolpx{} against \scasp{} and other Hybrid ASP solvers; utilization of \clingo{}'s multi-shot solving for more efficiency; improving the syntax  of \HEC{} (especially nested tuples); and finally, application of \HEC{} on planning problems in the general purpose definition language PDDL. 

\secVspaceB{}\section*{Acknowledgments}\secVspaceA{}

The work was supported by:
project 23-06506S of the Czech Science Foundation;
Horizon Europe project VASSAL (101160022);
project COSASS (PID2021-123673OB-C32) funded by MCIN/AEI/10.13039/501100011033; 
project EVASAI (PID2024-158227NB-C32) funded by MICIU/AEI/10.13039/501100011033 and FEDER, EU;
a research gift from Nexco Corp;
and grant DFG SCHA 550/15, Germany.



\bibliographystyle{tlplike}
\bibliography{bibliography,scasp}

\appendix{}

\renewcommand\thesection{Appendix~\Alph{section}}
\renewcommand\thesubsection{\Alph{section}.\arabic{subsection}}

\renewcommand{\thetable}{\Alph{section}\arabic{table}}

\renewcommand{\thetable}{\Alph{section}\arabic{table}}

\section{Additional Details for Sections from the Paper}
\label{appendix:a}
This appendix contains further details which did not fit into the paper.
In particular, a~full version of a~table of experiments (instead of a~shortened version) in Section~\ref{apx:sec:comp-expr-dec},
graphs of memory scaling in Section~\ref{apx:sec:memory-scaling}, 
a detailed graph of execution time scaling for \clingo{} in Section~\ref{apx:sec:full-scaling-clingo},
a full encoding of the counter example in Section~\ref{apx:example:counter},
a full encoding of the falling object example in Section~\ref{apx:example-falling-object},
short guidelines on transforming narratives from \DEC{} to \HEC{} in Section~\ref{sec:guid-transf-narr},
a full encoding of the \DEC{} axioms in Section~\ref{apx:sec:dec-axioms},
and a full encoding of the \HEC{} axioms in Section~\ref{apx:sec:hec-axioms}.

\newpage
\subsection{Comparing the expressiveness of \DEC{} and \HEC{}}
\label{apx:sec:comp-expr-dec}

\begin{table}[h]
  \setlength{\tabcolsep}{4pt}
  \linespread{0.9}
  \centering
  \smaller
  \caption{Comparing the expressiveness of \DEC{} and \HEC{}.} 
  \label{fig:tb-comparison-expressiveness-full}
  \begin{tabular}{p{5.2cm}lccccc}
    \toprule                                                                                                                                                                      
    \multirow{2}{*}{Reasoning Type and Notable Features} & \multirow{2}{*}{Problem}                                                                                                           & \DEC{}~\linkIconDecAx{}     & \multicolumn{2}{c}{\HEC{}~\linkIconHecAx{}} \\
                                                         &                                                                                                                                    &     \clingo{} &      \clingcon{} &      \clingolpx{} \\
    \midrule                                             
    D: event effect                                      & \linkIconDecEx{ex1-light_on_off}~\linkIconHecEx{ex1-light_on_off}                                         Light on/off             & \yes       & \yes          & \yes            \\  %
    \gmidrule                                            
    D: state constraint                                  & \linkIconDecExECASP{DeadOrAlive40-ea.lp}~\linkIconHecExECASP{DeadOrAlive40-ea.lp}                         DeadOrAlive              & \yes       & \yes          & \yes            \\  
                                                         & \linkIconDecExECASP{Happy40-ea.lp}~\linkIconHecExECASP{Happy40-ea.lp}                                     Happy                    & \yes       & \yes          & \yes            \\  
                                                         & \linkIconDecExECASP{StuffyRoom40-ea.lp}~\linkIconHecExECASP{StuffyRoom40-ea.lp}                           StuffyRoom               & \yes       & \yes          & \yes            \\  
                                                         & \linkIconDecEx{ex13-carrying_a_book/model-state_constr.lp}~\linkIconHecEx{ex13-carrying_a_book/model-state_constr.lp}                         Carrying a book 2        & \yes       & \yes          & \yes            \\  %
    \gmidrule                                            
    D: conditional effect axiom                          & \linkIconDecExECASP{Telephone40-ea.lp}~\linkIconHecExECASP{Telephone40-ea.lp}                             Telephone                & \yes       & \yes          & \yes            \\  
                                                         & \linkIconDecExECASP{Yale40-ea.lp}~\linkIconHecExECASP{Yale40-ea.lp}                                       Yale                     & \yes       & \yes          & \yes            \\  
                                                         & \linkIconDecEx{ex5-light_toggle}~\linkIconHecEx{ex5-light_toggle}                                   Light toggle             & \yes       & \yes          & \yes            \\  %
    \gmidrule                                            
    D: effect constraint                                 & \linkIconDecExECASP{WalkingTurkey40-ea.lp}~\linkIconHecExECASP{WalkingTurkey40-ea.lp}                     WalkingTurkey            & \yes       & \yes          & \yes            \\  
                                                         & \linkIconDecEx{ex13-carrying_a_book/model-effect_constr.lp}~\linkIconHecEx{ex13-carrying_a_book/model-effect_constr.lp}                   Carrying a book 1        & \yes       & \yes          & \yes            \\  %
    \gmidrule                                            
    \multirow[t]{3}{5.2cm}{M: non-determinism            
    (release from inertia / determining fluent)}         & \linkIconDecExECASP{RussianTurkey40-ea.lp}~\linkIconHecExECASP{RussianTurkey40-ea.lp}                     RussianTurkey            & \yes       & \yes          & \yes            \\  
                                                         & \linkIconDecExECASP{ChessBoard40-ea.lp}~\linkIconHecExECASP{ChessBoard40-ea.lp}                           ChessBoard               & \yes       & \yes          & \yes            \\  
                                                         & \linkIconDecExECASP{CoinToss40-ea.lp}~\linkIconHecExECASP{CoinToss40-ea.lp}                               CoinToss                 & \yes       & \yes          & \yes            \\  
                                                         & \linkIconDecEx{ex12-dice_roll}~\linkIconHecEx{ex12-dice_roll}                                             Dice roll                & \yes       & \yes          & \yes            \\  %
    \gmidrule                                            
    D: event trigger axiom                               & \linkIconDecEx{ex2-bank_account_nofee}~\linkIconHecEx{ex2-bank_account_nofee}                             Simpl. Bank account      & \yes       & \yes          & \yes            \\  %
    \gmidrule                                            
    D: concurrent event                                  & \linkIconDecExECASP{Supermarket40-ea.lp}~\linkIconHecExECASP{Supermarket40-ea.lp}                         SuperMarket              & \yes       & \yes          & \yes            \\  
    \gmidrule                                            
    A: abducible event occurrence                        & \linkIconDecExECASP{StolenCar40-ea.lp}~\linkIconHecExECASP{StolenCar40-ea.lp}                             StolenCar                & \yes       & \yes          & \yes            \\  
    \gmidrule                                            
    A: abd. fluent, disjunctive event trigger            & \linkIconDecExECASP{BusRide40-ea.lp}~\linkIconHecExECASP{BusRide40-ea.lp}                                 BusRide                  & \yes       & \yes          & \yes            \\  
    \gmidrule                                            
    D: compound event                                    & \linkIconDecExECASP{Commuter15-ea.lp}~\linkIconHecExECASP{Commuter15-ea.lp}                               Commuter                 & \yes       & \yes          & \yes            \\  
    \gmidrule                                            
    D: causal constraint                                 & \linkIconDecExECASP{ThielscherCircuit40-ea.lp}~\linkIconHecExECASP{ThielscherCircuit40-ea.lp}             ThielscherCircuit        & \yes       & \yes          & \yes            \\  
    \gmidrule                                            
    D: functional fluent                                 & \linkIconDecEx{ex15-counter}~\linkIconHecEx{ex15-counter}                                                 Counter                  & \yes       & \yes          & \yes            \\  %
                                                         & \linkIconDecEx{ex6-adder}~\linkIconHecEx{ex6-adder}                                                       Adder                    & \yes       & \yes          & \yes            \\  %
    \gmidrule                                            
    \multirow[t]{3}{4.35cm}{D: gradual change (integer), 
    event trigger, release from inertia}                 & \linkIconDecExECASP{FallingObjectWithEvents40-ea.lp}~\linkIconHecExECASP{FallingObjectWithEvents40-ea.lp} FallingObject            & \yes       & \yes          & \yes            \\  
                                                         & \linkIconDecExECASP{KitchenSink_M40-ea.lp}~\linkIconHecExECASP{KitchenSink_M40-ea.lp}                     KitchenSink              & \yes       & \yes          & \yes            \\  
                                                         & \linkIconDecExECASP{HotAirBalloon40-ea.lp}~\linkIconHecExECASP{HotAirBalloon40-ea.lp}                     HotAirBalloon            & \yes       & \yes          & \yes            \\  
                                                         & \linkIconDecEx{ex7-pulsing_light}~\linkIconHecEx{ex7-pulsing_light}                                       Pulsing light            & \yes       & \yes          & \yes            \\  %
                                                         & \linkIconDecEx{ex14-collision}~\linkIconHecEx{ex14-collision}                                             Collision                & \yes       & \yes          & \yes            \\  %
    \gmidrule                                            
    \multirow[t]{3}{4.5cm}{D: continuous change          
    (rational), event trigger, release from inertia}     & \linkIconDecEx{ex3-falling_object}~\linkIconHecEx{ex3-falling_object}                                     FallingObject            & \no        & \no           & \yes            \\  %
                                                         & \linkIconDecExECASP{KitchenSink_M40-ea.lp}~\linkIconHecExECASP{KitchenSink_M40-ea.lp}                     KitchenSink              & \no        & \no           & \yes            \\  
                                                         & \linkIconDecExECASP{HotAirBalloon40-ea.lp}~\linkIconHecExECASP{HotAirBalloon40-ea.lp}                     HotAirBalloon            & \no        & \no           & \yes            \\  
                                                         & \linkIconDecEx{ex7-pulsing_light}~\linkIconHecEx{ex7-pulsing_light}                                       Pulsing light            & \no        & \no           & \yes            \\  %
                                                         & \linkIconDecEx{ex14-collision}~\linkIconHecEx{ex14-collision}                                             Collision                & \no        & \no           & \yes            \\  %
    \gmidrule                                            
    D: Zeno-like behavior                                & \linkIconDecEx{ex4-bank_account}~\linkIconHecEx{ex4-bank_account}                                         Bank account             & \yesbut    & \yesbut       & \yes            \\  %
    \gmidrule                                            
    D: trivial Zeno behavior                             & \linkIconDecEx{ex10-blinking_light}~\linkIconHecEx{ex10-blinking_light}                                   Blinking light           & \yesbut    & \yesbut       & \yes            \\  %
    \gmidrule                                            
    D: true Zeno behavior                                & \linkIconDecEx{ex8-bouncing_ball}~\linkIconHecEx{ex8-bouncing_ball}                                       Bouncing ball            & n/a        & n/a           & \yes            \\  %
                                                         & \linkIconDecEx{ex9-water_tanks}~\linkIconHecEx{ex9-water_tanks}                                           Water tanks              & n/a        & n/a           & \yes            \\  %
    \bottomrule\\[-.5em]
\multicolumn{5}{c}{\small  Reasoning types: D=deduction, M=model finding, A=abduction.}
  \end{tabular}
\end{table}

\newpage
\subsection{Memory Scaling Graphs}
\label{apx:sec:memory-scaling}
\captionsetup[subfigure]{labelformat=mysubfig}
\begin{figure}[h!]
   \includegraphics[width=0.65\textwidth]{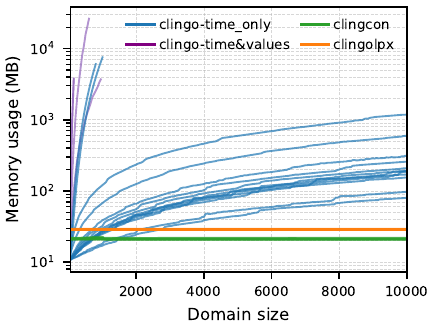}
    \caption{Memory usage scaling with the size of the domain.}
    \label{apx:fig:comparison-execution-domain-mem}
\end{figure}
\begin{figure}[h!]
    \includegraphics[width=0.65\textwidth]{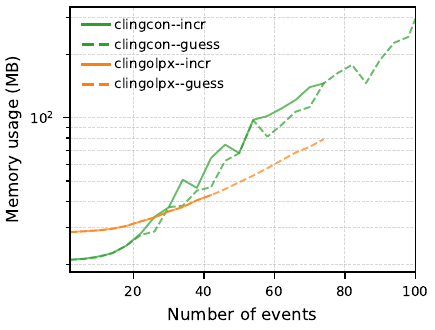}
    \caption{Memory usage scaling with the number of events.}
    \label{apx:fig:comparison-execution-falling-mem}
\end{figure}
\captionsetup[subfigure]{labelformat=parens}

\newpage
\subsection{Detailed Scaling Graphs for Clingo}
\label{apx:sec:full-scaling-clingo}
\begin{figure}[h!]
   \includegraphics[width=\textwidth]{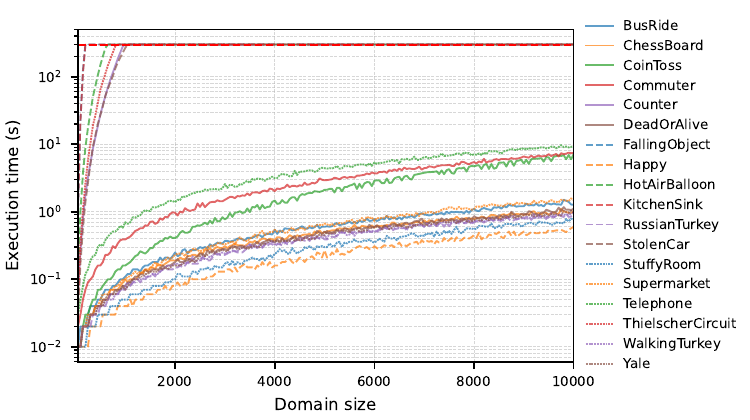}
    \caption{Execution time scaling with the size of the domain.}
    \label{fig:comparison-execution-domain-full-clingo}
\end{figure}
\begin{figure}[h!]
   \includegraphics[width=\textwidth]{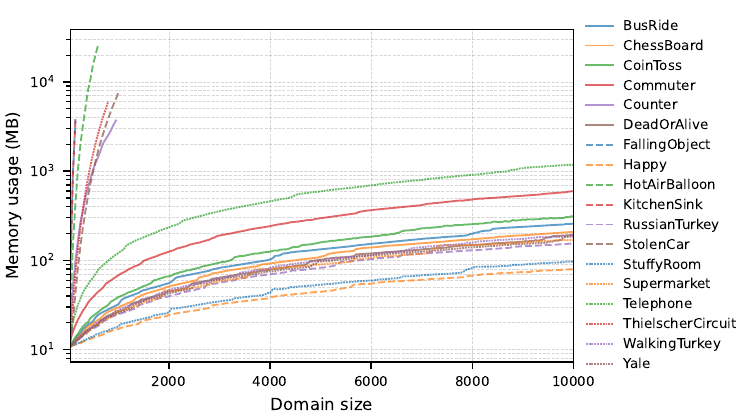}
    \caption{Memory usage scaling with the size of the domain.}
    \label{fig:comparison-execution-falling-full-clingo}
\end{figure}

\newpage
\subsection{Example: The Counter}
\label{apx:example:counter}

\begin{example}[The Counter~\linkIconDecEx{ex15-counter}~\linkIconHecEx{ex15-counter}]
  Figure~\ref{apx:fig:ex2-counter-full} shows the \DEC{} and \HEC{} encodings for the example of a counter by~\cite{mueller-ec-chapter} which can be incremented or reset. Notice that under Hybrid EC a definition of a domain for the fluent is not needed (line~1) and that terminates is not needed, since the old value is inherently replaced by the new value.
\end{example}

\begin{figure}[h]
  \begin{subfigure}[t]{.52\textwidth}
    \begin{lstlisting}[style=MySCASP, basewidth=.42em]
fluent(val(V)) :- dom(V).      
dom(0..10).
initiates(inc,val(V+1),T) :- holdsAt(val(V),T).
terminates(inc,val(V),T) :- holdsAt(val(V),T).
initiates(rst,val(0),T) :- time(T).
terminates(rst,val(V),T) :- V!=0, dom(V),time(T).
    \end{lstlisting}    
    \sfigCapVspaceB{}\caption{Original encoding.}
    \label{sfig:ex2-counter-orig}
  \end{subfigure}
  \begin{subfigure}[t]{.47\textwidth}
    \begin{lstlisting}[style=MySCASP]
ffluent(val).
% INC: (val,S+1) = 1*(val,S) +  1*1    
initiates(inc,val,((1,(val,S)),(1,1)),S):- 
   step(S).
% RST: (val,S+1) = 0*0
initiates(rst,val,((0,0),),S) :- step(S).
    \end{lstlisting}
    \sfigCapVspaceB{}\caption{Hybrid encoding}
    \label{sfig:ex2-counter-hybrid}
  \end{subfigure}
  \caption[]{Encoding of the counter example by~\cite{mueller-ec-chapter}}
  \label{apx:fig:ex2-counter-full}
\end{figure}

\newpage
\subsection{Example: Falling Object}
\label{apx:example-falling-object}

\begin{example}[Falling Object~\linkIconDecEx{ex3-falling_object}~\linkIconHecEx{ex3-falling_object}]
  Consider that an object is dropped from some initial height, falls via a constant speed of 2 units, and eventually hits the ground at height zero.
  Encoding of this example for \DEC{} is shown in Figure~\ref{apx:sfig:ex3-falling-object-orig} and the corresponding \HEC{} encoding in Figure~\ref{apx:sfig:ex3-falling-object-hybrid}.
  Notice that we do not need to define the domain for the height or the constraint that ensures only one value at a time (lines 2-4).
  Dropping the object makes it start falling and releases its value from inertia (lines 6-9).
  The height of the object changes according to a trajectory (lines 11-13) while falling.
  And the object hits the ground at height zero, which stops it falling (lines 15-21/22). 
  Referencing the value of height at \code{T1/S1} is done inside of the linear expression instead of using \code{holdsAt} (line 12),
  and using the value of height as a condition is done via the \code{&sum} statement instead of \code{holdsAt} (line 17).
\end{example}

\begin{figure}[h]
  \begin{subfigure}[t]{.52\textwidth}
    \begin{lstlisting}[style=MySCASP, basewidth=.43em]
fluent(height(O,H)) :- object(O), fHeight(H).
fHeight(-20..20).
H1 = H2 :- holdsAt(height(O,H1), T),
           holdsAt(height(O,H2), T).
    
initiates(drop(A,O), falling(O), T) :-
  agent(A), object(O), time(T).
releases(drop(A,O), height(O,H), T) :-
  agent(A), object(O), fHeight(H), time(T).

trajectory(falling(O),T1,height(O,H2),T2) :-
  H2 = H-2*T2, holdsAt(height(O,H), T1), 
  time(T2), time(T1+T2).
    
happens(hitGround(O), T) :- 
  holdsAt(falling(O), T), 
  holdsAt(height(O,0) ,T).
terminates(hitGround(O), falling(O), T) :-
  object(O), time(T).
initiates(hitGround(O), height(O, H), T) :-
  holdsAt(height(O, H), T).
\end{lstlisting}
    \sfigCapVspaceB{}\caption{Original encoding.}
    \label{apx:sfig:ex3-falling-object-orig}
  \end{subfigure}
  \begin{subfigure}[t]{.47\textwidth}
    \begin{lstlisting}[style=MySCASP]
ffluent(height(O)) :- object(O).
% domain not needed
% constraint not needed
%   only one value possible at a time

initiates(drop(A,O), falling(O), S) :-
  agent(A), object(O), step(S).
releases(drop(A,O), height(O), S) :-
  agent(A), object(O), step(S).

trajectory(falling(O), S1, height(O),
        (((1,(height(O),S1)),),-2), S2) :-
  object(O),step(S1),step(S2),step(S1+S2).

happens(hitGround(O), S) :- 
  holdsAt(falling(O), S),
  &sum{ (height(O), S) } = 0, step(S).
terminates(hitGround(O), falling(O), S) :-
  object(O), step(S).
initiates(hitGround(O), height(O), 
          ((1,(height(O),S))), S) :-
  object(O), step(S).

no_jump(height(O), S, 0) :- object(O),
  holdsAt(falling(O), S), step(S).
    \end{lstlisting}
    \sfigCapVspaceB{}\caption{Hybrid encoding}
    \label{apx:sfig:ex3-falling-object-hybrid}
  \end{subfigure}
  \caption[]{Encoding of the falling object example by~\cite{mueller_book-fixed}}
  \label{fig:ex3-falling-object}
\end{figure}

\newpage
\subsection{Guidelines for Transforming From DEC to HEC}
\label{sec:guid-transf-narr}
In this section we provide more guidelines with examples on how to convert a problem (i.e., an example) from \DEC to \HEC.
We also provide over 30 examples encoded both in \DEC and \HEC at \repo{} and archived at \artifactDoi{}.
EC specifications of problem instances consist of a universal theory, a domain description, and a narrative.
The universal theory is provided by the \DEC or \HEC axioms.
The domain description includes the definition of fluents, events, and the rules of the domain.
And the narrative consists of observations of event occurrences and about fluents.

First, one must replace \DEC with \HEC, as provided in Section~\ref{apx:sec:hec-axioms}, and, since \HEC reasons about steps instead of time, all occurrences of \code{time/1} must be replaced with \code{step/1} in the domain description and in the narrative of the particular problem instance.
Then, the domain description and the narrative needs to be transformed, as discussed below.

All fluents in the original \DEC encoding of a domain description are relational.
One must consider which fluents should be made into functional fluents.
In general, propositional/boolean fluents should stay relational and fluents with numerical arguments should be converted into functional fluents, unless their value domain is small enough to not cause too much grounding.
If a fluent is ever subject to a (anti-)trajectory then it must be represented as a functional fluent. 
There are no modifications needed for fluents that remain relational.

\parVspaceB{}\paragraph{Converting holdsAt:} To convert a relational fluent with a numeric argument, e.g., \code{height(O, X)}, into a functional fluent, its declaration must be changed from \code{fluent(height(O, X))} to \code{ffluent(height(O))}, shifting its value argument
into the theory part of the reasoning.
Only one dimension of a fluent can be converted directly and only if exactly one of its possible values holds at any given timepoint. 
Any occurrences of \code{holdsAt(height(O,X), T), X = LE} for the new functional fluent must be replaced with a \code!&sum{LE} = (height(O), S)! statement, where \code{LE} is a linear expression and other relational operators can appear instead of the equality.
Examples of this transformation can be found in Figure~\ref{apx:fig:transforming-holdsat}.
Notice that in the last example the value of the height of the basket has been moved into the \code{&sum} statement of the height of the apple.
This is because they must relate to each other within one linear equation.

Multi-dimensional fluents can either be converted partially, or they must first be represented using multiple one-dimensional fluents, e.g., \code{position(X,Y)} can be split into \code{positionX(X)} and \code{positionY(Y)}.
Fluents for which multiple values can hold at the same timepoint cannot be converted, although, it should be possible to convert them to N functional fluents for up to N values at the same time.
Fluents for which no value holds at some timepoint must be adjusted to always have a value.
This can be achieved by assigning a dedicated value for the cases where there was no value, e.g., zero, and by introducing a new relational fluent which holds at the timepoints where its associated functional fluent has a valid value.
If the value of the new functional fluent is referenced in the body of a rule, then also introduce \code{holdsAt} or \code{not holdsAt} into the rule for the new, associated relational fluent.
For example, check that a calculator is \code{powered_on} before checking the \code{result} shown on its display.

\begin{figure}[tb]
\figVspaceB{} 
  \begin{subfigure}[t]{.49\textwidth}
    \begin{lstlisting}[style=MySCASP]
% in the body
head :- holdsAt(height(apple, 0), T).
% in a constraint (headless rule)
:- holdsAt(height(apple, X), T), 0 < X,
   holdsAt(onTheFloor(apple), T).
% in the head with a ground value
holdsAt(height(apple, 0), T) :-
  holdsAt(onTheFloor(apple), T).
% in the head with another func. fluent
holdsAt(height(apple, X), T) :-
  holdsAt(height(basket, X), T),
  holdsAt(isIn(apple, basket), T).
    \end{lstlisting}    
    \sfigCapVspaceB{}\caption{Original encoding.}
    \label{sfig:transforming-holdsat-orig}
  \end{subfigure}
  \begin{subfigure}[t]{.49\textwidth}
    \begin{lstlisting}[style=MySCASP]

head :- &sum{ 0 } = (height(apple), S).

:- &sum{ 0 } < (height(apple), S),
   holdsAt(onTheFloor(apple), S).

&sum{ 0 } = (height(apple), S) :-
  holdsAt(onTheFloor(apple), S).

&sum{ (height(basket), S) }
  = (height(apple), S) :-
  holdsAt(isIn(apple, basket), S).
    \end{lstlisting}
    \sfigCapVspaceB{}\caption{Hybrid encoding}
    \label{sfig:transforming-holdsat-hybrid}
  \end{subfigure}
  \caption{Transforming \code{holdsAt/2} to \code{&sum/1}.}
  \label{apx:fig:transforming-holdsat}
\end{figure}


\parVspaceB{}\paragraph{Converting initiates, terminates, releases:} Occurrences of \code{initiates(E,height(O,X),T), X = LE} must be replaced with \code{initiates(E,height(O),LE,S)}, where \code{LE} is a nested tuple representing a linear expression that defines the value to be initiated, as was defined in Section~\ref{sec:functional-fluents} of the paper.
Examples of this transformation can be found in Figure~\ref{apx:fig:transforming-initiates}.
\begin{figure}[tb]
  \begin{subfigure}[t]{.49\textwidth}
    \begin{lstlisting}[style=MySCASP]
% incrementing the current value by 1
initiates(lift_up(O), height(O, X+1), T) :-
  holdsAt(height(O, X), T).
% combine the values of two fluents
initiates(place_on(O1,O2),
          height(O1,X1+X2),T) :-
  holdsAt(height(O1, X1), T),
  holdsAt(height(O2, X2), T).
    \end{lstlisting}    
    \sfigCapVspaceB{}\caption{Original encoding.}
    \label{sfig:transforming-initiates-orig}
  \end{subfigure}
  \begin{subfigure}[t]{.49\textwidth}
    \begin{lstlisting}[style=MySCASP]

initiates(lift_up(O), height(O),
  ((1,(height(O),S)), (1,1)), S) :- step(S).

initiates(place_on(O1, O2),
          height(O1),
  ( (1,(height(O1),S)),
    (1,(height(O2),S)) ), S) :- step(S).
    \end{lstlisting}
    \sfigCapVspaceB{}\caption{Hybrid encoding}
    \label{sfig:transforming-initiates-hybrid}
  \end{subfigure}
    \caption{Transforming \code{initiates/3} to \code{initiates/4}.}
  \label{apx:fig:transforming-initiates}
\end{figure}
%
%
Occurrences of \code{terminates(E,height(O,X),T)} that are coupled with an \code{initiates/3}, i.e., terminating the old value while initiating a new one, are no longer needed. 
If the occurrence is not coupled with an \code{initiates/3}, then its effect causes the fluent to have no value.
Thus, it should be replaced by initiating a zero and changing the value of a helper fluent that represents the lack of value of the fluent affected by the terminates.
%
Occurrences of \code{releases(E,height(O,X),T)} are replaced by \code{releases(E,height(O),S)}, and occurrences of \code{releasedAt(height(O,X),T)} by \code{releasedAt(height(O),S)}. However, note that since functional fluents can only have one value at a time, it is not possible to release just some values of the fluent. Either all possible values of the fluent are released or none are.

Functional fluents are used in the following examples provided in our \github{}: \linkHecExample{ex2-bank\_account\_nofee}{Bank account (no fee)}, \linkHecExample{ex3-falling\_object}{Falling object}, \linkHecExample{ex4-bank\_account}{Bank account}, \linkHecExample{ex6-adder}{Adder}, \linkHecExample{ex7-pulsing_light}{Pulsing light}, \linkHecExample{ex8-bouncing_ball}{Bouncing ball}, \linkHecExample{ex9-water_tanks}{Water tanks}, \linkHecExample{ex12-dice_roll}{Dice roll}, \linkHecExample{ex13-carrying_a_book}{Carrying a book}, \linkHecExample{ex14-collision}{Collision}, \linkHecExample{ex15-counter}{Counter}, \linkHecExampleECASP{FallingObjectWithAntiTrajectory25-ea.lp}{FallingObjectWithAntiTrajectory}, \linkHecExampleECASP{FallingObjectWithEvents40-ea.lp}{FallingObjectWithEvents}, \linkHecExampleECASP{HotAirBalloon40-ea.lp}{HotAirBalloon}, and \linkHecExampleECASP{KitchenSink_M40-ea.lp}{KitchenSink}.

\parVspaceB{}\paragraph{Converting trajectories/antiTrajectories:} Occurrences of \code{trajectory(F1,T1,F2,T2)} need to be replaced by \code{trajectory(F1,S1,F2,(LE,Rate),S2)}, where \code{LE} is a nested tuple representing a linear expression that defines the value of \code{F2} at \code{S1} that will change with time via rate \code{Rate}, as was defined in Section~\ref{sec:functional-fluents} of the paper.
An example of this can be seen in Section~\ref{apx:example-falling-object}.

Continuous change is used in the following examples provided in our \github{}: \linkHecExample{ex3-falling\_object}{Falling object}, \linkHecExample{ex7-pulsing_light}{Pulsing light}, \linkHecExample{ex8-bouncing_ball}{Bouncing ball}, \linkHecExample{ex9-water_tanks}{Water tanks}, \linkHecExample{ex13-carrying_a_book}{Carrying a book}, \linkHecExample{ex14-collision}{Collision}, \linkHecExampleECASP{FallingObjectWithAntiTrajectory25-ea.lp}{FallingObjectWithAntiTrajectory}, \linkHecExampleECASP{FallingObjectWithEvents40-ea.lp}{FallingObjectWithEvents}, \linkHecExampleECASP{HotAirBalloon40-ea.lp}{HotAirBalloon}, and \linkHecExampleECASP{KitchenSink_M40-ea.lp}{KitchenSink}.

\parVspaceB{}\paragraph{Converting triggered events:} Triggered events need \code{no_jump} constraints, as was discussed in Section~\ref{sec:mapping-steps} of the paper. The constraint should mirror the trigger condition of the event.
An example of this can be seen in Section~\ref{apx:example-falling-object}, where the trigger condition requires the object to be falling and the height to be zero.
In that case the \code{no_jump} constraint prevents jumps over height zero for all steps at which the object is falling.
In some cases, e.g., when triggering solely based on a relational fluent or based on some minimum duration, a custom manual constraint is needed instead of just the \code{no_jump} predicate, examples that feature these are marked using a * below.

Triggered events are used in the following examples provided in our \github{}: \linkHecExample{ex3-falling\_object}{Falling object}, \linkHecExample{ex4-bank\_account}{Bank account}*, \linkHecExample{ex7-pulsing_light}{Pulsing light}, \linkHecExample{ex8-bouncing_ball}{Bouncing ball}, \linkHecExample{ex9-water_tanks}{Water tanks}*, \linkHecExample{ex10-blinking_light}{Blinking light}*, \linkHecExample{ex13-carrying_a_book}{Carrying a book}, \linkHecExample{ex14-collision}{Collision}, \linkHecExampleECASP{BusRide40-ea.lp}{BusRide}, \linkHecExampleECASP{FallingObjectWithAntiTrajectory25-ea.lp}{FallingObjectWithAntiTrajectory}, \linkHecExampleECASP{FallingObjectWithEvents40-ea.lp}{FallingObjectWithEvents}, \linkHecExampleECASP{HotAirBalloon40-ea.lp}{HotAirBalloon} and \linkHecExampleECASP{KitchenSink_M40-ea.lp}{KitchenSink}.

\parVspaceB{}\paragraph{Converting observations in a narrative:}
All observations must be transformed to use the \code{obs/3} predicate, as described in Section~\ref{sec:modeling-narrative} of the paper.
Alternatively, we also use \code{initiallyP} and \code{initiallyN} as a shorthand to specify observations at time zero.

Observations are used in all of the examples in our \github{}.

\newpage
\subsection{DEC Axioms}\label{apx:sec:dec-axioms}

\begin{lstlisting}[style=MySCASP]
time(0..maxtime).
{holdsAt(F,T)} :- fluent(F), time(T).      % left open (i.e., released
{releasedAt(F,T)} :- fluent(F), time(T).   % from stability checking)

% DEC1
stoppedIn(T1,F,T2) :-
  happens(E,T),
  T1<T, T<T2,
  terminates(E,F,T),
  event(E), fluent(F), time(T), time(T1), time(T2).

% DEC2
startedIn(T1,F,T2) :-
  happens(E,T),
  T1<T, T<T2,
  initiates(E,F,T),
  event(E), fluent(F), time(T), time(T1), time(T2).

% DEC3
holdsAt(F2,T1+T2) :-
  happens(E,T1),
  initiates(E,F1,T1),
  0<T2,
  trajectory(F1,T1,F2,T2),
  not stoppedIn(T1,F1,T1+T2),
  event(E), fluent(F1), fluent(F2), time(T1), time(T1+T2).

% DEC4
holdsAt(F2,T1+T2) :-
  happens(E,T1),
  terminates(E,F1,T1),
  0<T2,
  antiTrajectory(F1,T1,F2,T2),
  not startedIn(T1,F1,T1+T2),
  event(E), fluent(F1), fluent(F2), time(T1), time(T1+T2).

% DEC5
holdsAt(F,T+1) :-
  holdsAt(F,T),
  not releasedAt(F,T+1),
  not terminated1(F,T),
  fluent(F), time(T), T<maxtime.

% DEC6
:- holdsAt(F,T+1),
  not holdsAt(F,T),
  not releasedAt(F,T+1),
  not initiated1(F,T),
  fluent(F), time(T), T<maxtime.

% DEC7
releasedAt(F,T+1) :-
  releasedAt(F,T),
  not initiated1(F,T),
  not terminated1(F,T),
  fluent(F), time(T), T<maxtime.

% DEC8
:- releasedAt(F,T+1),
  not releasedAt(F,T),
  not released1(F,T),
  fluent(F), time(T), T<maxtime.

% DEC9
holdsAt(F,T+1) :-
  happens(E,T),
  initiates(E,F,T),
  event(E), fluent(F), time(T), T<maxtime.

% DEC10
:- holdsAt(F,T+1),
  happens(E,T),
  terminates(E,F,T),
  event(E), fluent(F), time(T), T<maxtime.

% DEC11
releasedAt(F,T+1) :-
  happens(E,T),
  releases(E,F,T),
  event(E), fluent(F), time(T), T<maxtime.

% DEC12
:- releasedAt(F,T+1),
  happens(E,T),
  initiates(E,F,T),
  event(E), fluent(F), time(T), T<maxtime.
:- releasedAt(F,T+1),
  happens(E,T),
  terminates(E,F,T),
  event(E), fluent(F), time(T), T<maxtime.

% Auxiliary predicates
initiated1(F,T) :-
  happens(E,T),
  initiates(E,F,T),
  event(E), fluent(F), time(T).
terminated1(F,T) :-
  happens(E,T),
  terminates(E,F,T),
  event(E), fluent(F), time(T).
released1(F,T) :-
  happens(E,T),
  releases(E,F,T),
  event(E), fluent(F), time(T).
\end{lstlisting}

\newpage
\subsection{HEC Axioms}
\label{apx:sec:hec-axioms}

\begin{lstlisting}[style=MySCASP]
step(0..laststep).
{    holdsAt(F,S) } :-  fluent(F), step(S).
{ releasedAt(F,S) } :-  fluent(F), step(S).
{ releasedAt(F,S) } :- ffluent(F), step(S). 
% value of variables in linear equations is open by default

% HEC1 only for relational fluents (identical to DEC1)
stoppedIn(S1,F,S2) :-
  S1<S, S<S2,
  happens(E,S),
  terminates(E,F,S),
  step(S1), step(S2).

% HEC2 for both types of fluents (added a clause with initiates/4)
startedIn(S1,F,S2) :-
  S1<S, S<S2,
  happens(E,S),
  initiates(E,F,S),
  step(S1), step(S2).
startedIn(S1,F,S2) :-
  S1<S, S<S2,
  happens(E,S),
  initiates(E,F, _, S),
  step(S1), step(S2).

% HEC3 only for functional fluents (replaced DEC3)
&sum{ C*V : @member(LE) = (C,V); R*(time,S1+S2); -R*(time,S1) } = (F2,S1+S2) :-
  happens(E,S1),
  initiates(E,F1,S1),
  0<S2,
  trajectory(F1,S1,F2,(LE,R),S2),
  not stoppedIn(S1,F1,S1+S2),
  step(S1+S2).

% HEC4 only for functional fluents (replaced DEC4)
&sum{ C*V : @member(LE) = (C,V); R*(time,S1+S2); -R*(time,S1) } = (F2,S1+S2) :-
  happens(E,S1),
  terminates(E,F1,S1),
  0<S2,
  antiTrajectory(F1,S1,F2,(LE,R),S2),
  not startedIn(S1,F1,S1+S2),
  step(S1+S2).

% HEC5 for both types of fluents (added a clause without holdsAt
% and initiated instead of terminated)
holdsAt(F,S+1) :-
  holdsAt(F,S),
  not releasedAt(F,S+1),
  not terminated1(F,S),
  step(S+1).
&sum{ (F,S1) } = (F,S1+1) :-
  not releasedAt(F,S1+1), 
  not initiated1(F,S1),
  ffluent(F), step(S1), step(S1+1).

% HEC6 only for relational fluents (identical to DEC6)
:- holdsAt(F,S+1),
  not holdsAt(F,S),
  not releasedAt(F,S+1),
  not initiated1(F,S),
  step(S), step(S+1).

% HEC7 for both types of fluents (identical to DEC7)
releasedAt(F,S+1) :-
  releasedAt(F,S),
  not initiated1(F,S),
  not terminated1(F,S),
  step(S+1).

% HEC8 for both types of fluents (identical to DEC7)
:- releasedAt(F,S+1),
  not releasedAt(F,S),
  not released1(F,S),
  step(S), step(S+1).

% HEC9 for both types of fluents (added a clause with a theory atom)
holdsAt(F,S+1) :-
  happens(E,S),
  initiates(E,F,S),
  step(S+1).
&sum{ C*V : @member(LE) = (C,V) } = (F,S1+1) :-
  happens(E,S1),
  initiates(E,F,LE,S1),
  step(S1+1).

% HEC10 only for relational fluents (identical to DEC10)
:- holdsAt(F,S+1),
  happens(E,S),
  terminates(E,F,S),
  step(S+1).

% HEC11 for both types of fluents (identical to DEC11)
releasedAt(F,S+1) :-
  happens(E,S),
  releases(E,F,S),
  step(S+1).

% HEC12 for both types of fluents (added a clause with initiates/4)
:- releasedAt(F,S+1),
  happens(E,S),
  initiates(E,F,S),
  step(S+1).
:- releasedAt(F,S+1),
  happens(E,S),
  terminates(E,F,S),
  step(S+1).
:- releasedAt(F,S+1),
  happens(E,S),
  initiates(E,F,_,S),
  step(S+1).

% Auxiliary predicates
initiated1(F,S)  :- happens(E,S), initiates(E,F,S).
initiated1(F,S)  :- happens(E,S), initiates(E,F,_,S).
terminated1(F,S) :- happens(E,S), terminates(E,F,S).
% no terminated1(F,S) for ffluent(F)
released1(F,S)   :- happens(E,S), releases(E,F,S).
% released1(F,S) is defined above given releases(E,F,S) for ffluent(F)

% external python function member(t) for clingcon
% defined as member(_,t) for clingo-lpx due to a newer version of python
#script (python)
def member(t):
  return t.arguments 
#end.
\end{lstlisting}

\end{document}